\documentclass[aps, prl, a4paper, showpacs, twocolumn, superscriptaddress,10pt]{revtex4-1}

\usepackage{textcomp,hyperref,enumerate, url, cancel}                    			
\usepackage[margin=0.8in]{geometry}
\usepackage[monochrome]{color}
\usepackage{amsmath,amssymb,amsthm,bbm,bm,nicefrac}   				
\usepackage{graphicx,epsfig}                                                               		
\usepackage{multirow}

\newcommand{\ssection}[1]{\emph{#1.---}}

\newcommand{\ket}[1]{\left|#1\right\rangle}							
\newcommand{\bra}[1]{\left\langle#1\right|}
\newcommand{\proj}[1]{\ket{#1}\!\!\bra{#1}}
\newcommand{\ketbra}[2]{\left|#1\rangle\langle#2\right|}
\newcommand{\braket}[2]{\left\langle #1\lvert#2\right\rangle}
\newcommand{\abs}[1]{\left\lvert #1\right\rvert}

\newcommand{\dketbra}[1]{\left|#1\right\rangle\left\langle#1\right|}

\newcommand{\ui}{\underline{i}}
\newcommand{\uj}{\underline{j}}

\newcommand{\be}{\begin{equation}} 							
\newcommand{\ee}{\end{equation}}
\newcommand{\ba}{\begin{align}}
\newcommand{\ea}{\end{align}}
\newcommand{\bematrix}{\left(\begin{matrix}}
\newcommand{\ematrix}{\end{matrix}\right)}

\theoremstyle{definition}

\theoremstyle{theorem}

\newtheorem*{theorem*}{Theorem}

\theoremstyle{lemma}

\theoremstyle{proposition}

\theoremstyle{corollary}

\theoremstyle{observation}

\theoremstyle{remark}

\def\one{{\mbox{$1 \hspace{-1.0mm}  {\bf l}$}}}

\newcommand{\tr}[1]{\mathrm{Tr}\left[#1\right]}
\newcommand{\ptr}[2]{\mathrm{Tr}_{#1}\left[#2\right]}
\def\d{\mathrm{d}}

\def\cB{\mathcal B}

\def\cD{\mathcal D}
\def\cE{\mathcal E}

\def\cG{\mathcal G}
\def\cH{\mathcal H}											
\def\cI{\mathcal I}

\usepackage[normalem]{ulem}

\begin{document}
\title{Beyond the swap test: optimal estimation of quantum state overlap}
\author{M. Fanizza}\email{marco.fanizza@sns.it}
\affiliation{NEST, Scuola Normale Superiore and Istituto Nanoscienze-CNR, I-56126 Pisa, Italy}
\author{ M. Rosati}\email{Matteo.Rosati@uab.cat}
\affiliation{F\'isica Te\`orica: Informaci\'o i Fen\`omens Qu\`antics, Departament de F\'isica, Universitat Aut\`onoma de 
Barcelona, 08193 Bellaterra (Barcelona) Spain}
\author{M. Skotiniotis}
\affiliation{F\'isica Te\`orica: Informaci\'o i Fen\`omens Qu\`antics, Departament de F\'isica, Universitat Aut\`onoma de 
Barcelona, 08193 Bellaterra (Barcelona) Spain}
\author{J. Calsamiglia}
\affiliation{F\'isica Te\`orica: Informaci\'o i Fen\`omens Qu\`antics, Departament de F\'isica, Universitat Aut\`onoma de 
Barcelona, 08193 Bellaterra (Barcelona) Spain}
\author{V. Giovannetti}
\affiliation{NEST, Scuola Normale Superiore and Istituto Nanoscienze-CNR, I-56126 Pisa, Italy}
\begin{abstract}
We study the estimation of the overlap between two unknown pure quantum states of a finite dimensional system, given $M$ and $N$ copies of each type. This is a fundamental primitive in quantum information processing that is commonly accomplished from the outcomes of $N$ swap-tests, a joint measurement on one copy of each type whose outcome probability is a linear function of the squared overlap.  We show that a more precise estimate can be obtained by allowing for general collective measurements on all copies. 
We derive the statistics of the optimal measurement and compute the optimal mean square error in the asymptotic pointwise and finite Bayesian estimation settings. Besides, we consider two strategies relying on the estimation of one or both the states, and show that, although they are suboptimal, they outperform the swap test. In particular, the swap test is extremely inefficient for small values of the overlap, which become exponentially more likely as the dimension increases. 
Finally, we show that the optimal measurement is less invasive than the swap test and study the robustness to depolarizing noise for qubit states. 
\end{abstract}
\maketitle

\ssection{Introduction}
The overlap between two unknown quantum states is an archetypical instance of quantum relative information~\cite{OverlapLSB,Unspeakable,ReviewFramesInfo,ResourceAsymmetry,AsymmetryBasic,AsymmetryModes, genSchurEst, ChiribellaMo} and the estimation of the overlap is a basic primitive in quantum information processing,
with applications ranging from quantum fingerprinting~\cite{SwapTest,QFinger1,QFinger2}, entanglement estimation~\cite{FuncOverlap,EntDet1,EntDet2,EntDet3}, communication without a shared reference frame~\cite{OverlapBRS, OverlapBIM, OverlapLSB, OverlapGI} to quantum machine learning~\cite{HHL,LloydAlgo, QuantumSVM,ProgrammableProjective,LearningSwap,GaussianProc,NNAlgorithm,IBMSupervised,Nana,variationalnear}. 
Recently, with the advent of quantum machine learning protocols~\cite{HHL,LloydAlgo, QuantumSVM,GaussianProc,NNAlgorithm,IBMSupervised,Nana}, overlap estimation (OvE)  has attracted renewed interest as a fundamental primitive and its efficient implementation and generalization on near-term quantum computers are subjects of current research~\cite{ProgrammableProjective,LearningSwap}.
In most applications, OvE is carried out through the swap test (SWT)~\cite{SwapTest, LearningSwap, ProgrammableProjective}:
given two systems in the state $\ket{\psi}\ket{\phi}$, the probability of projecting it on its symmetric or antisymmetric part is determined by the overlap between $\ket{\psi}$ and $\ket{\phi}$. By repeating this measurement on several pairs of copies one can obtain a good estimate of this probability, and hence of the overlap.
It is then natural to ask whether, for the same number of copies, one could reach a larger accuracy via a collective strategy that extracts the relevant information using a joint and less-destructive measurement. In this article we answer in the positive, evaluating the ultimate precision attainable in the OvE of two pure quantum states, given a number of copies of each and assuming no prior knowledge about them.

The task we consider is as follows: 
given $N$ and $M\geq N$ copies of two unknown pure states $\ket{\psi},\ket{\phi}$ of a $d$-dimensional quantum system 
we are requested to provide an estimate of their (squared) overlap $\lvert\braket{\psi}{\phi}\rvert^2$ which is fixed, but unknown to us. The task is carried out by a machine that performs a measurement on the state
$\ket{\Psi}=\ket{\psi}^{\otimes N}\otimes\ket{\phi}^{\otimes M}$ of $M+N$ qudits and produces an estimate with maximum precision, as quantified by the mean square error (MSE).
Furthermore we consider the case of unlabeled states, i.e., when the machine receives $U_{\sigma}\ket{\Psi}$, with $U_\sigma$ an unknown permutation of the qudits. Note that in this case OvE constitutes in itself an instance of unsupervised quantum-classical learning problem, in a setting similar to \cite{SentisUnsuper}.

The measurements optimizing the average information gain~\cite{OverlapBRS} and the average error~\cite{OverlapBIM,OverlapLSB} have been derived for the case of qubits, with only numerical solutions~\cite{OverlapGI} for higher dimensions.
\begin{figure}[t!]
\includegraphics[scale=.28]{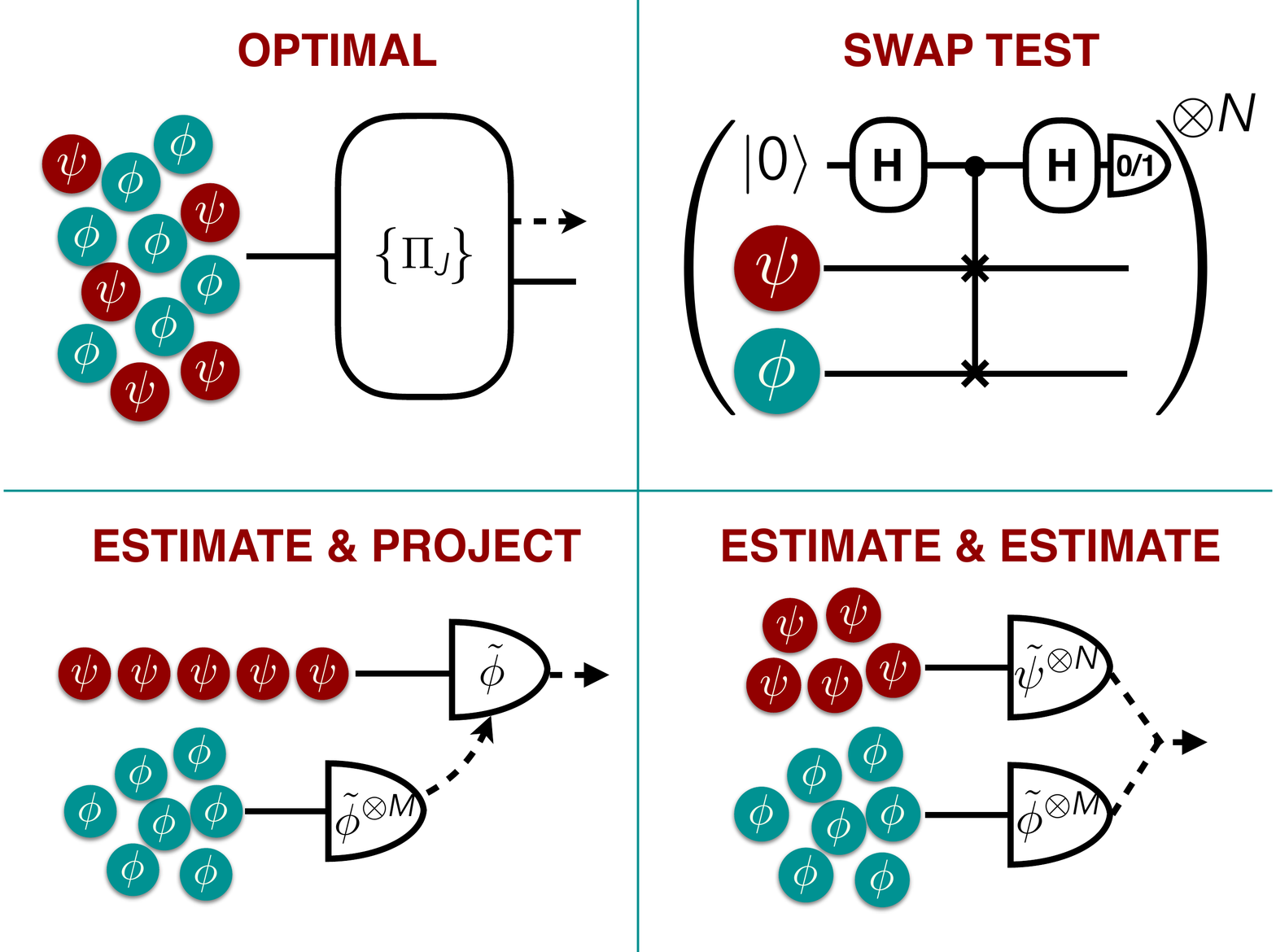}
\caption{Sketch of the OvE strategies studied in the article. a) Optimal measurement, e.g. by Schur transform (see Ref.~\cite{Schur3} for the circuit implementation). b) Circuit for the SWT, to be repeated $N$ times. c) Estimate $\ket{\phi}$ and project $\ket{\psi}$ on the estimated direction. d) Estimate both  $\ket{\phi}$ and 
$\ket{\psi}$ and calculate the overlap.}
\label{fig0}
\end{figure}
Here we tackle OvE in full generality, characterizing the optimal estimation within both local (pointwise) and global (Bayesian) approaches. For local estimation, we provide an asymptotically achievable lower bound using the quantum Fisher information (QFI)~\cite{ HelstromBOOK, FisherParis}, whereas for Bayesian estimation~\cite{HelstromBOOK,Personick} we provide an exact solution, generalizing the results of~\cite{OverlapGI}. We find that the optimal local strategy is also Bayesian-optimal asymptotically and that it performs identically in the labeled and unlabeled scenarios.
 We compare our results with the SWT and with two LOCC strategies based on estimating either one or both $\ket{\psi}$ and $\ket{\phi}$, see Fig.~\ref{fig0}. Such strategies are useful in distributed scenarios where copies of $\ket{\psi}$ and $\ket{\phi}$ are produced in different and distant laboratories. We show that in the limit of large $M+N$ and $|M-N|$ constant the optimal strategy displays a finite asymptotic gap with respect to all the others. Moreover, we show that the optimal measurement is less invasive than the SWT and robust against single-qubit noise. 
 
\ssection{Assessing the machine's performance}\label{sec:measures}
The states $\ket\psi=U\ket0$ and $\ket\phi=V\ket0$ are drawn uniformly at random, i.e., with Haar-distributed $U,V\in SU(d)$. Upon performing a measurement $\{E_{k}\}$ on $\ket\Psi$ with outcome $k$, the machine outputs an estimate $c(k)$ of the overlap $c=\lvert\braket{\psi}{\phi}\rvert^2$, with squared error $(c(k)-c)^{2}$.

In the \emph{global approach}, the machine's performance is quantified by averaging the squared error over all possible states and outcomes:
\be\label{eq:global_var}
v=\sum_{k}\int dU\,dV\,(c(k)-c)^{2} \tr{E_{k}\proj{\Psi}}.
\ee 
We refer to $v$ as \textit{global} MSE. Writing $V=U W$ and using the Haar-measure invariance $d V= d W$ and $c=|\bra{0}W\ket{0}|^{2}$ the average mean square error can be written as: \mbox{$v=\sum_{k}\int dW\,(c(k)-c)^{2} \tr{E_{k}\rho(c)}$}, where we have defined the effective state
\be
\rho(c)=\int\d U \, U^{\otimes(N+M)}\, \proj{\Psi_{0}}\, U^{\dagger \otimes(N+M)}.\label{eq:averageState}
\ee
where $\ket{\Psi_{0}}=\one\otimes W^{\otimes M} \ket{0}^{\otimes (N+M)}$.
In addition, we can write the  above integral over $W$ as an integral over the overlap  such that 
\be
v=\sum_{k}\int dc\; p(c) \,(c(k)-c)^{2} \tr{E_{k}\rho(c)}
\ee
where the distribution over overlaps is given by~\cite{OverlapStatistics,SentisUnsuper}
\be
 p(c)=\int dU \delta (c-|\bra{0}U\ket{0}|^2)=(d-1)(1-c)^{d-2}.
\label{eq:prior}
\ee
From the above discussion we see that the average over both types of states, i.e. over $U$ and $V$, is equivalent to an average over overlaps with weight $p(c)$ and over different orientations, $U^{\otimes(N+M)}$. This is a direct consequence  of the fact that if the states are completely unknown, then all pairs of states with equal overlap are equally probable and are related by a rigid unitary: 
 $\lvert\braket{\psi}{\phi}\rvert^2=\lvert\braket{\psi'}{\phi}'\rvert^2$ if and only if it exists $U$ such that $\ket{\psi'}=U\ket{\psi}$ and $\ket{\phi'}=U \ket{\phi}$.

{\color{red} At variance with the global approach, where the overlap is a random variable, in the \emph{local approach} the  overlap is considered to be fixed}. We then assess the performance of the machine by computing the average of the square error over all states with fixed overlap $c$ and over all outcomes:
\be\label{eq:local_var}
v(c)=\sum_{k} (c(k)-c)^{2} \tr{E_{k}\rho(c)},
\ee 
also in terms of the average state for a fixed overlap $\rho(c)$. We refer to $v(c)$ as \textit{local} MSE. 
As shown in the supplemental material (SM), {\color{red}the integral in Eq.~(\ref{eq:averageState}) can be performed using $SU(d)$ representation theory and $SU(2)$ Clebsch-Gordan coefficients, obtaining the block-diagonal form}
\be
\rho(c)=\sum_{J=J_{\min}}^{J_{\max}}p(J|c)\frac{\one_J}{\chi_{J}}\otimes\dketbra{\sigma}_{J},
\label{eq:avgstate}
\ee
with $J_{\min}=\frac{\abs{M-N}}{2}$, $J_{\max}=\frac{M+N}{2}$ and
\be
\small p(J|c)=\frac{(2J+1)N!M!(1-c)^{M}P_{J+J_{\min}}^{(0,-2J_{\min})}\left(\frac{1+c}{1-c}\right)}{(J_{\max}-J)!(J_{\max}+1+J)!},
\label{eq:prob_dist}
\ee
where $P_n^{(\alpha,\beta)}(x)$ is the n$^\mathrm{th}$-degree Jacobi polynomial. In the previous equations, for $d=2$, $J$ is the familiar total-angular-momentum label and $\one_{J}=\sum_{M=-J}^J\proj{J,M}$ is the projector on the subspace of total angular momentum $J$, of dimension $\chi_{J}=2J+1$. In general, for $d>2$,
  $\one_{J}$ are projectors over the subspaces of dimension $\chi_J(d)$ hosting irreducible representations (irreps) of $\mathrm{SU}(d)$ arising from the tensor product of two completely symmetric representations of $M$ and $N$ qudits; these irreps are still indexed by an (half)-integer $J\in[J_{\min},J_{\max}]$. Finally, $\ket{\sigma}_{J}$ is a state representing the known labeling of the states and it belongs to the irrep-space of the permutation group, also labeled by $J$. Note that, in the unlabeled scenario, the average over qudit permutations acts only on $\ket{\sigma}_{J}$ for each $J$, depolarizing it to a projector on the whole irrep space. Importantly, note that all the information about the overlap is contained in the $J$-statistics $p(J|c)$, which is independent of dimension and labeling. In particular, the optimal measurement is given by the projectors $\Pi_J$ on the subspaces labeled by $J$, and it can be implemented via weak Schur sampling~\cite{Schur3}. Indeed, for any POVM $\{E_k\}_{k}$, we can get the same outcome probability distribution if we use the POVM $\{\Pi_J E_k\Pi_J\}_{k,J}$ and then ignore the $J$ label. {\color{red}When this POVM is applied to $\rho(c)$ the outcome probabilities are $p_{k,J}:=p(k|J)p(J|c)$. The same outcome probabilities $p_{k,J}$ can be generated by applying directly the POVM $\{\Pi_J\}_J$ followed by classical post-processing. The latter can only increase the variance of the estimator, by convexity of the figure of merit: $\sum_k p(k|J)(c_{k,J}-c)^2\geq( \sum_k p(k|J) c_{k,J}-c)^2$, which follows from the Cauchy-Schwartz inequality. Therefore $\{\Pi_J\}_J$ is optimal both for local and global estimation and one can replace $\tr{E_{J}\rho(c)}$ with $p(J|c)$ in Eqs.~(\ref{eq:global_var},\ref{eq:local_var}), effectively reducing our problem to one of classical estimation, i.e., optimizing the function $c(J)$.}
  
\ssection{Local estimation}
\label{sec:fisherinfo}
The classical Cramer-Rao 
bound~\cite{HolevoBook} places a lower bound on the MSE of all local unbiased estimators $c(J)$ as 
$
v(c)\geq H(c)^{-1}
$,       
where 
$
H(c) =\sum_{J}(\partial_{c}p(J|c))^{2}/p(J|c)$ is the Fisher information of the measurement statistics.
In the limit $M+N\to\infty$ and $M-N\ll (M+N)\sqrt{c}$, we can use an approximation of the Jacobi polynomial given in \cite{Szego1939} to obtain the following asymptotically-unbiased estimator and its associated MSE:
\be\label{eq:QFI_overlap}\small
c_{\rm op}^{\rm loc}(J)=\left(\frac{2J}{M+N}\right)^2, \quad 
v_{\rm op}(c)=\frac{4c(1-c)}{M+N},
\ee
In the SM we show that $v_{\rm op}(c)$ coincides with $H(c)^{-1}$ to leading order in $\frac{1}{M+N}$ and hence the Cramer-Rao bound is achievable in this limit.
If instead $M\rightarrow\infty$ and $N$ is finite, it is clear that $\ket\phi$ can be estimated perfectly and hence the optimal strategy is to project the copies of $\ket \psi$ in this known direction, with resulting $v_{\rm op}(c)=\frac{c(1-c)}{N}$.

\ssection{Bayesian estimation}
\label{sec:Bayesian}
The optimal classical Bayesian (global) estimator is given~\cite{HelstromBOOK} by $c_{\rm op}^{\rm bay}(J)=\frac{\int dc\,c\,p(c) p(J|c)}{\int dc\,p(c) p(J|c)}$. Using graphical calculus techniques for the recoupling theory of Clebsch-Gordan coefficients \cite{Varshalovich1988}, as explained in the SM, we obtain the following optimal global estimator and corresponding MSE:
{\small\begin{align}
c_{\rm op}^{\rm bay}(J)&=\frac{d+J+J^{2}+\frac{M+N}{2}-\left(\frac{M+N}{2}\right)^{2}+MN}{(d+M)(d+N)},\\
v_{\rm op}&=\frac{(d-1) (d + M + N)}{d (d+1) (d + M) (d + N)}.
\label{eq:opAvVar}
\end{align}}
We pause to highlight the following facts: i) when $d$ is fixed and the number of copies is large, the prior distribution of the states is little informative with respect to the information that can be obtained by the actual measurement; indeed we can see that when $M+ N\rightarrow \infty$, $M- N$ constant, $c_{\rm op}^{\rm bay}(J)\approx c_{\rm op}^{\rm loc}(J)$ implying that the local optimal estimator is also a good Bayesian estimator and viceversa; ii) contrarily to the local estimation results, the global MSE of Eq.~\eqref{eq:opAvVar} is exact for all $M$, $N$ and depends on $d$ due to the prior, Eq.~\eqref{eq:prior};  iii) in particular, $v_{\rm op}$ decays as $d^{-2}$ if one of either $M$ or $N$ is kept finite, whereas it decays only as $d^{-1}$ when $M, N\gg1$. 
\begin{table}[t!]
\begin{center}
\begin{tabular}{ |l|c|c|c| } 
 \hline
 \bf Local est. & ${\bf  v_{\rm op}(c)}$ & ${\bf  v_{\rm ep}(c)}$ & ${\bf  v_{\rm ee}(c)}$ \\ 
 \hline\hline
 $M=N\rightarrow \infty$ & $\frac{4c(1-c)}{M+N}$ & $\frac{3}{2}v_{\rm op}(c)$ & $2 v_{\rm op}(c)$  \\ 
 \hline
 $M\rightarrow\infty$ & $\frac{c(1-c)}{N}$ & $v_{\rm op}(c)$ & $2v_{\rm op}(c)$  \\ 
 \hline\hline
 \bf  Bayesian est. & $\bf  v_{\rm op}$ & $\bf  v_{\rm ep}$ & $\bf  v_{\rm ee}$ \\ 
 \hline\hline
  $M=N\rightarrow \infty$ & $\frac{4(d-1)}{d(d+1)(M+N)}$ & $\frac{3}{2}v_{\rm op}$ & $2 v_{\rm op}$ \\
  \hline
  $M\rightarrow\infty$ & $\frac{(d-1)}{d(d+1)(d+N)}$ & $v_{\rm op}$ & $\frac{d+2N}{2+N} v_{\rm op}$\\
  \hline
\end{tabular}
\end{center}
\caption{Local MSE and global MSE  attainable via the optimal, EP and EE strategies in two asymptotic limits. In all the cases 
the global MSEs coincide with the corresponding average local MSE values, apart from asymptotically vanishing corrections.}\label{tab1}
\end{table}

\ssection{1-LOCC strategies}
\label{sec:estiandproj}
We now consider a family of intermediate strategies that employ 1-LOCC on $\ket{\psi}^{\otimes N}$ and $\ket{\phi}^{\otimes M}$.
{\color{red}The estimate-and-project (EP) strategy consists in estimating $\ket{\phi}$ from its $M$ copies, then projecting each copy of $\ket{\psi}$ on this estimate and counting the fraction of successful projections. When $\ket{\phi}$ is known, projecting $\ket{\psi}$ on $\ket{\phi}$ is optimal \cite{HolevoBook}. {\color{red} However, EP is not necessarily the optimal 1-LOCC strategy.} The corresponding POVM elements can be written as $
E_{V,k}^{\rm(ep)}=dV E_{V}^{(M)}\otimes V^{\otimes N}\Pi_{k}^{(N)}V^{\dag \otimes N}$, where $E_{V}^{(M)}=\chi_{\frac{M}{2}}(d) (V\dketbra{0}V^{\dag})^{\otimes M}$ is the optimal covariant measurement to estimate $\ket\phi$  \cite{Hayashi1997, HayashiGroupRep} and $\Pi_{k}^{(N)}$ represents $k$ successful projections of the copies of $\ket{\psi}$ on the estimate of $\ket{\phi}$. The estimator is $c_{\rm ep}^{\rm loc}(k)=\frac kN$.}
The estimate-and-estimate (EE) strategy instead consists in estimating both $\ket{\psi}$ and $\ket{\phi}$ separately, then computing the overlap between the estimated states. The corresponding POVM elements can be written as $E_{V,W}^{\rm(ee)}=dV dW E_{V}^{(M)}\otimes E_{W}^{(N)}$, i.e., a product of two covariant measurements to estimate $\ket{\phi}$ and $\ket{\psi}$. We take as local estimator $c_{\rm ee}^{\rm loc}(V,W)=\abs{\bra{0}V^{\dag}W\ket{0}}^{2}$.
In the SM we provide exact results for local and Bayesian estimation using EP and EE. Table~\eqref{tab1} compares the performance of these strategies with the optimal one in two asymptotic limits. We find that, for both local and Bayesian estimation, the EE strategy is always worse than the optimal by a factor of $2$, whereas the EP strategy attains a MSE equal to the optimal in the limit $M\rightarrow\infty$, $N$ finite.

\begin{figure}[t!]
\includegraphics[scale=.45]{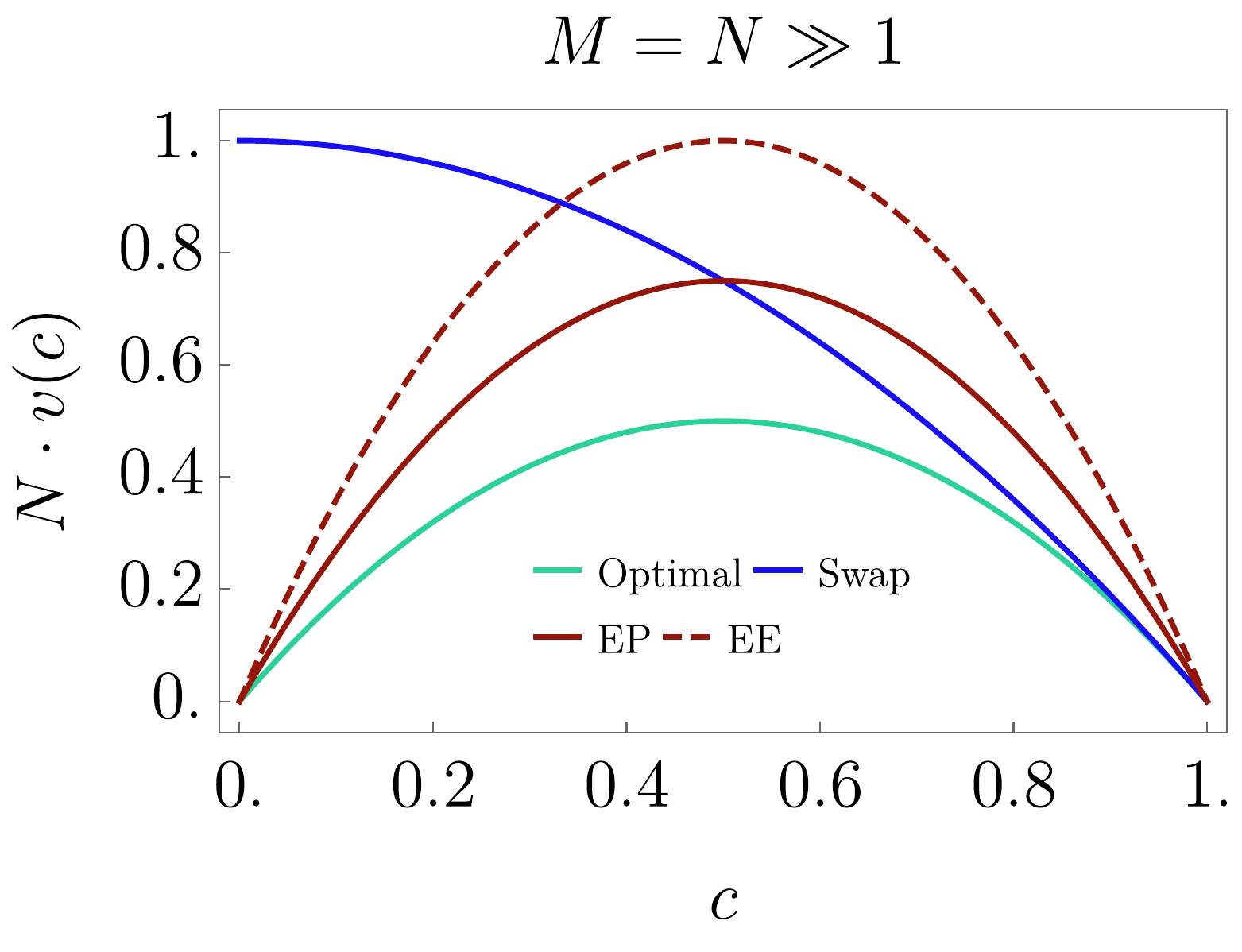}
\caption{Plot of the optimal local MSE scaling coefficient $N\cdot v(c)$ vs.~the true value of the overlap $c$, at leading order in $M=N$, for the strategies analyzed in the article.} \label{fig1}
\end{figure}

\ssection{Performances comparison} We now compare the strategies discussed so far with the traditional SWT~\cite{SwapTest}. Note that all these strategies except the optimal one require labeling of the states.
The latter consists in projecting the state $\ket{\psi}\otimes\ket{\phi}$ on its triplet/singlet components, hence it coincides with the optimal measurement for $M=N=1$. As the SWT acts on couples of states, we restrict to the case $M=N$. The probability of a triplet projection $p(c)=\frac{1}{2}\left(1+c\right)$ and the ensuing statistics of $k$ successful projections out of $N$ trials is given by the binomial distribution. The optimal local MSE attainable by this test is well-known, $v_{\rm sw}(c)=\frac{1-c^{2}}{N},\label{Varswap}$ while for the optimal global MSE $v_{\rm sw}$ one can derive an exact expression for each value of $k$, then compute the sum numerically, as detailed in the SM. In the asymptotic limit of $M=N\gg d$ a good approximation is provided by averaging the optimal local MSE:  $v_{\rm sw} \simeq \int dc\;p(c) v_{\rm sw}(c)=(d+2)(d-1)/(d(d+1)N)$, which is $\sim d$ times larger than $v_{\rm op}$.

In the same limit, we can compare the local MSE of all the strategies, see Fig.~\ref{fig1}. First, we observe a gap between the optimal strategy, that attains the QFI, and all the other strategies. This means that, even with a large number of copies, the collective measurement on $\ket\psi^{\otimes N}\otimes\ket \phi^{\otimes M}$ has a clear advantage over a non-collective one. Second, we observe that the relative error $\frac{\sqrt{v(c)}}{c}$ for small $c$ scales as $\frac{1}{c \sqrt{N}}$ for the SWT and as $\frac{1}{\sqrt{cN}}$ for the other strategies, implying a quadratic improvement in $\frac 1 {\sqrt{c}}$ in the number copies needed to reach a fixed relative error, while the optimal measurement is still computationally efficient (see next section). This is particularly relevant since for large $d$ small overlaps are exponentially more likely, see Eq. \ref{eq:prior}. This phenomenon is also at the source of the so-called ``barren plateau" problem~\cite{googleplateau, glassyplateau} for quantum variational circuits, and other types of strategies have been proposed to address this issue \cite{plateaulearning,quantumplateau,unitaryoverlap}. 

\begin{figure*}[t!]
\includegraphics[scale=.35]{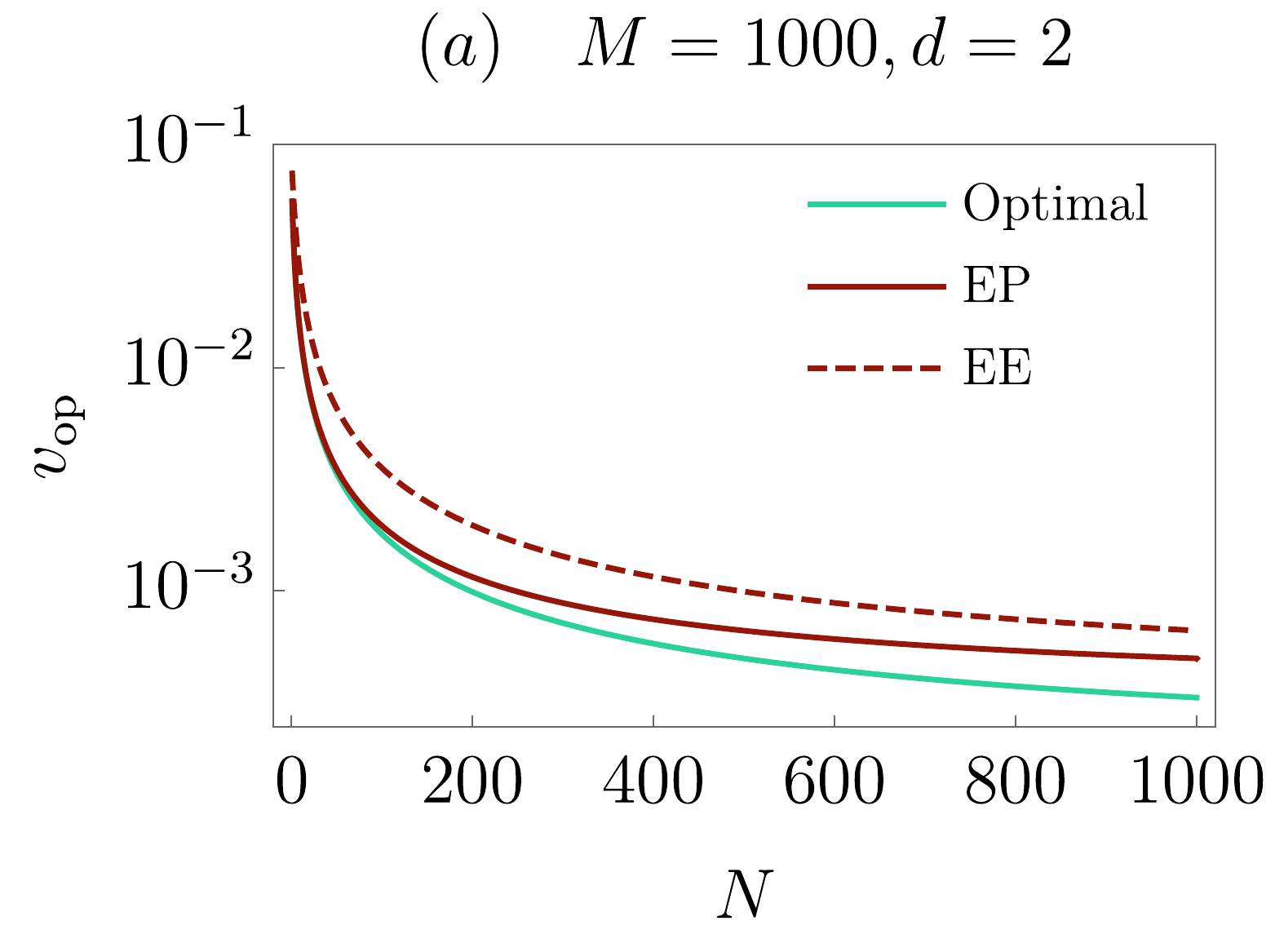}\includegraphics[scale=.35]{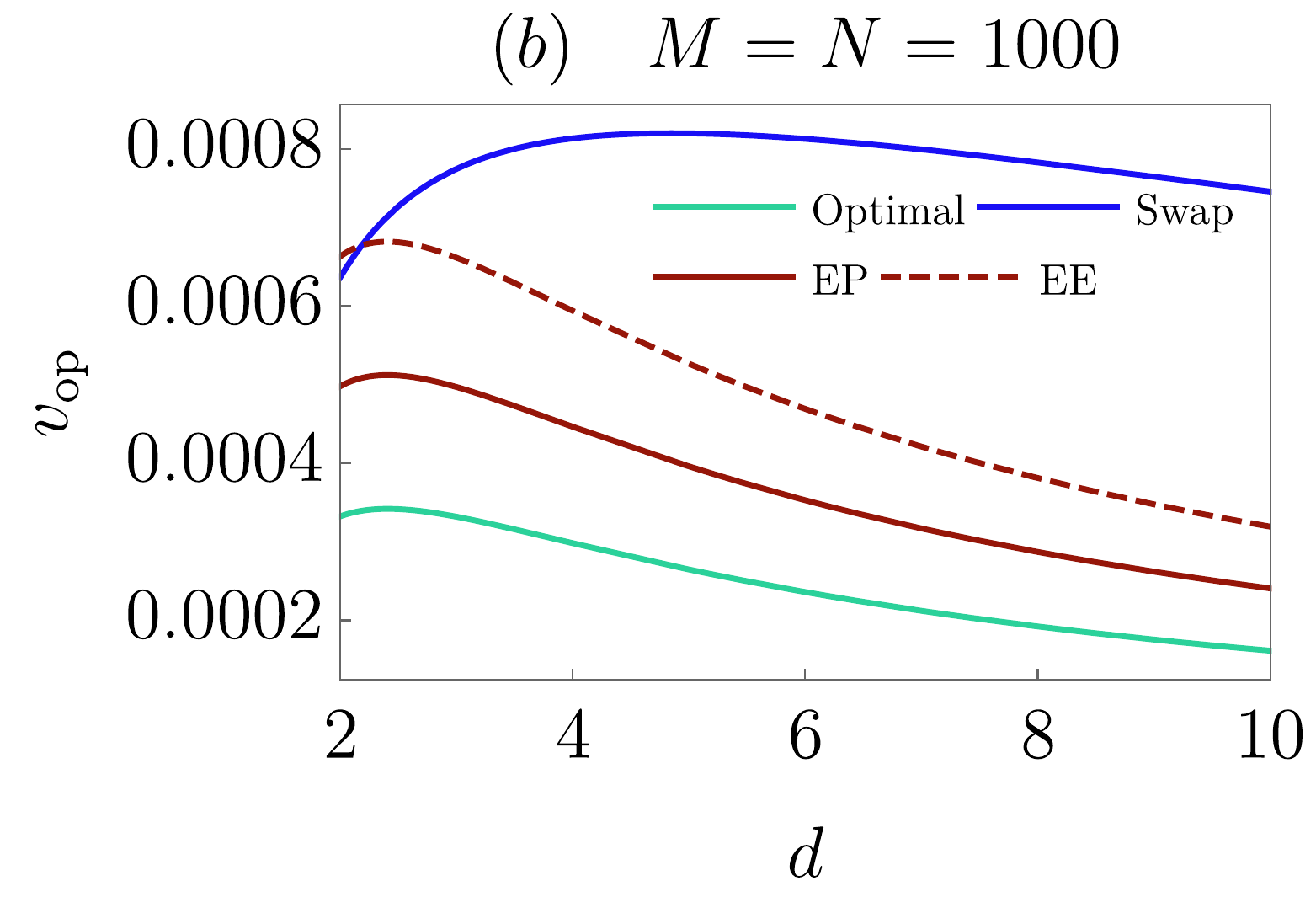}\includegraphics[scale=.35]{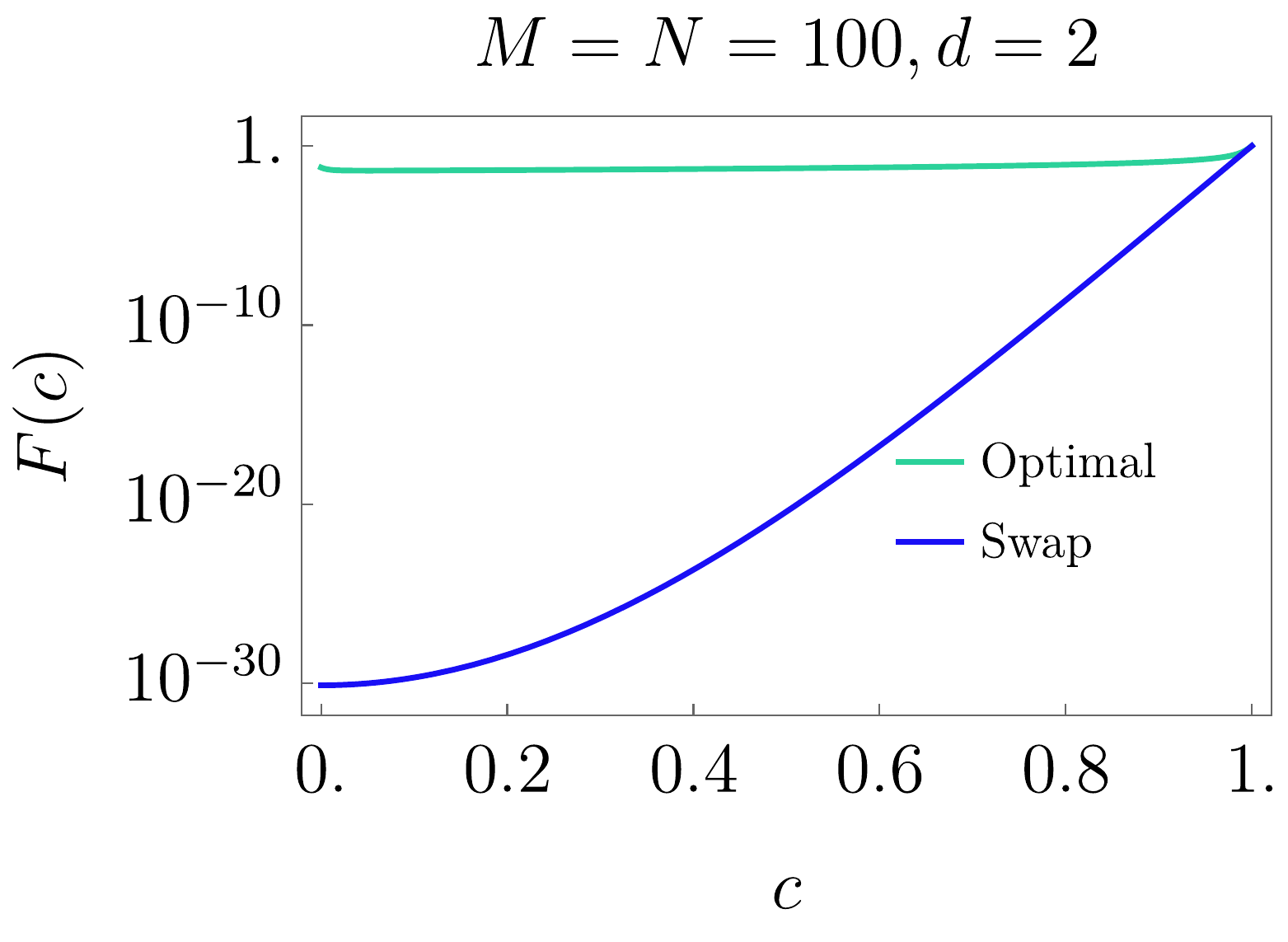}
\caption{
(a) Plot of the optimal global MSE $v_{opt}$ vs.~the number of copies of one state $N$, for a fixed number of copies of the other $M=1000$, in dimension $d=2$, for the optimal, EP and EE strategies. (b) Plot of the optimal global MSE $v_{opt}$ vs.~the dimension $d$, for a fixed number of copies $M=N=1000$ for all the strategies studied. (c) Plot of the average post-measurement fidelity with the initial state $F(c)$ vs.~the true value of the overlap $c$ with a fixed and equal number of copies $M=N=100$, for the optimal strategy and SWT.} \label{fig2}
\end{figure*}

We notice similar features for the global MSE, plotted in Fig.~\ref{fig2} as a function of $N$ for $M$ fixed and increasing $d$ (inset). We observe that the SWT is comparable with EE for $M\sim N$ and $d=2$, but with a small increase in dimension this feature disappears. Moreover, there is in general a gap between the EP and EE strategies, the former being closer to the optimal one.

\ssection{Gate complexity} The advantage in the precision of the optimal estimation comes with the tradeoff that the optimal measurement requires entangling operations over the whole system of $N+M$ qudits. The Schur transform ~\cite{Schur1,Schur2,Schur3} is a way to perform the optimal measurement, and requires $O(\mathrm{poly}(N+M, \log d, \log \frac 1 \epsilon))$ qudit gates for precision $\epsilon$. The resulting algorithm is efficient, but still unfeasible without error correction. The SWT instead requires $N$ independent circuits of fixed depth, and may still be convenient for large overlaps or very noisy gates.

A mid-term solution is to divide input data in $R$ groups of $S$ copies of $\ket{\phi}$ and $\ket{\psi}$, such that $S$ is the largest integer for which the given architecture can perform the optimal measurement with high fidelity, repeat the measurement $R$ times and do classical post-processing. The performances of these intermediate protocols are between SWTs and optimal measurement. See the SM for a more detailed discussion of these issues.

\ssection{Measurement invasiveness}
Another relevant figure of merit for applications is the fidelity between the post-measurement state and the initial one, averaged over the measurement outcomes.
Both the optimal measurement and the SWT are projective measurements. We assume that the post-measurement states are given by the result of such projections and hence the average post-measurement fidelity can be written as
\be\label{pfm}
F(c)=\int_{U\in \mathrm{SU}(d)}dU\sum_{k}\abs{\bra{\Psi_{U}}E_{k}\ket{\Psi_{U}}}^{2},
\ee
with $\{E_{k}\equiv\one_{J}\}$ for the optimal measurement and $\{E_{k}=\cG_{{\cal S}_{N}}(\one_{2}^{S\,\otimes k}\otimes\one_{2}^{A\,\otimes N-k})\}$ for the SWT, where $\one^{S/A}_{2}$ are the projectors on the singlet/triplet components of $\cH_{2}^{\otimes 2}$. Then Eq.~\eqref{pfm} is simply given by
\be\small
F_{\rm op}(c)=\sum_{J=J_{\min}}^{J_{\max}}p(J|c)^{2},~F_{\rm sw}(c)=\left(\frac{1+c^{2}}{2}\right)^{N},\label{eq:fidelity}
\ee 
as shown in the SM. In Fig.~\ref{fig2} we plot these two quantities as a function of $c$, showing that the optimal measurement is less invasive than the SWT, especially for small overlap values. 

\ssection{Noise-robustness}
Finally, we consider how the optimal strategy changes when the states, which are expected to be pure, are affected by depolarizing noise acting independently on each qudit before reaching the measurement stage. Note that if the noisy channel is of a different kind, one can at least reach the optimal MSE for the depolarizing channel by performing a twirling operation, realizable by pre- and post-processing with random unitaries on each qudit plus classical forward communication. This operation is $\int dU U^{\dag}{\cal N}(U\rho U^{\dag})U=\Delta_{r}(\rho)$ for some $r$, where we have defined the depolarizing channel as $\Delta_{r}=r\mathcal{I}+(1-r)\frac{\one}{d}\mathrm{Tr}$ and $\mathcal{I}$ is the identity channel. After this operation the overall state of the system can now be written as $\Delta_{r_{0}}(\psi)^{\otimes N}\otimes\Delta_{r_{1}}(\phi)^{\otimes M}$. 

In the SM we compute the optimal MSE in this case, restricting to $d=2$ for simplicity. In the limit $M,N\rightarrow\infty$ with $\frac MN$ finite, the global MSE at leading order is 
$
v_{\rm op,mix}=\frac{1}{6Mr_0^2}+\frac{1}{6Nr_1^2} 
$,
which agrees with the previously found limit of Eq.~\eqref{eq:opAvVar} for zero-noise, $r_{i}=1$. Hence the net effect of white noise is to rescale the MSE by a factor $r_{i}^{-2}$ for each state.

\ssection{Conclusions}
In this article we have computed the ultimate precision attainable in estimating the overlap of two arbitrary pure quantum states, as a function of the dimension of their Hilbert space and their number of copies. We showed that the commonly-used SWT is highly inefficient for small values of the overlap and also on average over Haar-distributed random states. The optimal strategy is a collective measurement on all the copies and can be implemented efficiently using the Schur transform, although it remains experimentally challenging. A practical alternative is to do Schur sampling on subsets of the dataset, followed by classical post-processing.
In addition, we proposed two intuitive strategies that estimate separately one or both states and showed that they also outperform the SWT. Finally, we showed that the optimal measurement is less invasive than the SWT and robust to white noise. The strategies we introduced provide several clear advantages over the SWT, and they could become a standard tool for various quantum technologies, as well as providing improvements in the runtime of quantum algorithms.

\ssection{Note added after publication} 
We thank one of the referees of TQC 2020 for pointing out~\cite{BOW} to us, where the authors find a minimum variance unbiased estimator of the Hilbert-Schmidt distance of two unknown mixed states, and compute its variance. Our results, valid for more general estimators, complement their analysis in the case of pure states.

\ssection{Acknowledgments}
M.\,F. and V.\,G. acknowledge support from PRIN 2017 ``Taming complexity with quantum strategies". 
 M.\,R., M.\,S. and J.\,C. acknowledge support from the Spanish MINECO, project FIS2016-80681-P with the support of AEI/FEDER funds; the Generalitat de Catalunya, project CIRIT 2017-SGR-1127. M.\,R. also acknowledges partial financial support by the  Baidu-UAB collaborative project ``Learning of Quantum Hidden Markov Models".
M.\,S. also acknowledges support from the Spanish MINECO project IJCI-2015-24643.
M.\,F. thanks Matthias Christandl for helpful comments. M.\,S. and J.\,C. acknowledge useful discussion with Nana Liu. 

\bibliographystyle{apsrev4-1}

  \appendix
\begin{widetext}
\begin{center} \bf SUPPLEMENTAL MATERIAL\end{center}
\tableofcontents
\section{Schur-Weyl duality and the irreducible representations of $\mathrm{SU(d)}$ and $S_{N+M}$}
\label{app:preliminaries}

We begin by first reviewing some key concepts and techniques in group representation theory that we will frequently make use of throughout this Supplemental 
material.  The most important ingredient is Schur-Weyl duality~\cite{Weyl}.  Consider  the state space of $N+M$, $d$-dimensional systems, 
$\cH_d^{\otimes (N+M)}$.  This space carries the action of two different groups;  the special unitary group of $d\times d$ complex matrices, $\mathrm{SU}(d)$, 
and the permutation group of $N+M$ objects, $S_{N+M}$.  Specifically, the groups $\mathrm{SU}(d)$ and $S_{N+M}$ act on a basis $\{\ket{i_1} \otimes \ket{i_2} \otimes ... \otimes \ket{i_{N+M}}\}_{i_1,i_2....,i_{N+M}}$ of $\cH_d^{\otimes(N+M)}$ via unitary 
representations $u_{N+M}:\mathrm{SU}(d)\to\mathrm{U}(\cH_d^{\otimes(N+M)})$, and $s_{N+M}:S_{N+M}\to\mathrm{U}(\cH_d^{\otimes(N+M)})$ as follows
\begin{align}\nonumber
u_{N+M}(U)\ket{i_1} \otimes \ket{i_2} \otimes ... \otimes \ket{i_{N+M}}&=U^{\otimes(N+M)}\ket{i_1} \otimes \ket{i_2} \otimes ... \otimes \ket{i_{N+M}}\nonumber\\&=U\ket{i_1} \otimes U\ket{i_2} \otimes ... \otimes U\ket{i_{M+N}},\quad \forall U\in\mathrm{SU}(d)\\
s_{N+M}(\sigma)\ket{i_1} \otimes \ket{i_3} \otimes ... \otimes \ket{i_{N+M}}&=\ket{\sigma^{-1}({i_{1}}) }\otimes \ket{\sigma^{-1}({i_2})} \otimes ... \otimes \ket{\sigma^{-1}({i_{N+M}})},\quad \forall\sigma\in S_{N+M}.
\label{supp:groupactions}
\end{align}
Observe that $[U^{\otimes(N+M)},s(\sigma)]=0,\; \forall U\in\mathrm{SU}(d),\,\mathrm{and}\, \forall\sigma\in S_{N+M}$.  Schur-Weyl duality states that the 
total state space $\cH_d^{\otimes(N+M)}$ can be decomposed as
\be
\cH_d^{\otimes(N+M)}\cong\bigoplus_Y \,\mathcal U^{(Y)}(\mathrm{SU}(d))\otimes\mathcal U^{(Y)}(S_{N+M}),
\label{supp:SchurWeyl}
\ee
where $\mathcal U^{(Y)}(\mathrm{SU}(d))$ is the space of dimension $\chi_Y$ upon which the unitary irreducible representation (irrep)  $u^{(Y)}$  of $\mathrm{SU}(d)$ acts and $\mathcal U^{(Y)}(S_{N+M})$ is the space of 
dimension $\omega_Y$ on which the irrep  $s^{(Y)}$ of $S_{N+M}$ acts.  
The use of the congruence sign in Eq.~\eqref{supp:SchurWeyl}  indicates that this block decomposition is accomplished by a unitary transformation; 
in the case considered here this unitary is the Schur transform~\cite{Schur1,Schur2}.

The label $Y$ indexes the integer partitions of $N+M$ in at most $d$ parts written in descending order, pictorially represented by Young diagrams, where $N+M$ boxes are arranged into at most $d$ rows.  
For $d=2$ the block decomposition of 
Eq.~\eqref{supp:SchurWeyl} is the familiar decomposition of $N+M$ spin-$\nicefrac{1}{2}$ systems into the total angular momentum 
label $J$.  The latter is related 
to Young diagrams of at most two rows where the number of boxes in any of the two rows is given by 
$(\frac{N+M}{2}+J, \frac{N+M}{2}-J)$.  

With the help of Eq.~\eqref{supp:SchurWeyl} we can express the action of $u_{N+M}$ and $s_{M+N}$ as 
\begin{align} \nonumber
u_{N+M}(U)&=\bigoplus_Y u^{(Y)}(U)\otimes \one, \quad\forall U\in\mathrm{SU}(d),\\
s(\sigma)&=\bigoplus_Y \one\otimes s^{(Y)}(\sigma),\quad \forall  \sigma\in S_{N+M}.
\label{supp:block irreps}
\end{align}
where we have made implicit the fact that $ u^{(Y)}(U)$ acts on $\mathcal U^{(Y)}(\mathrm{SU}(d))$ and $s^{(Y)}(\sigma)$ acts on  $\mathcal U^{(Y)}(S_{N+M})$ respectively.  Given a state $\rho\in\cB(\cH_d^{\otimes(N+M)})$, and applying $u_{N+M}(U)$ with $U$ extracted from the Haar measure $\d U$
of $\mathrm{SU}(d)$ gives rise to the following completely positive, trace-preserving (CPTP) map 
\be
\cG_{\mathrm{SU}(d)}[\rho]=\int\, \d U\, U^{\otimes(N+M)}\, \rho\, U^{\dagger\,\otimes(N+M)}, 
\label{supp:su(d)twirl}
\ee
Similarly, applying $s(\sigma)$ chosen uniformly at random gives rise to the following CPTP map
\be
\cG_{S_{N+M}}[\rho]=\frac{1}{(N+M)!}\sum_{\sigma} s_{N+M}(\sigma)\, \rho\, s_{N+M}^\dagger(\sigma).
\label{supp:sntwirl}
\ee
By decomposing the representations $u_{N+M}$, and $s_{N+M}$ into their irreps as in Eq.~\eqref{supp:block irreps}, and making use of Schur's 
lemmas~\cite{Sternberg} Eqs~(\ref{supp:su(d)twirl},~\ref{supp:sntwirl}) can be conveniently written as~\cite{ReviewFramesInfo}
\begin{align}\nonumber
\cG_{\mathrm{SU}(d)}[\rho]&=\bigoplus_Y\left(\cD_{\mathcal U^{(Y)}(\mathrm{SU}(d))}\otimes\cI_{\mathcal U^{(Y)}(S_{N+M})}\right)[\Pi_Y\,\rho]\\
\cG_{S_{N+M}}[\rho]&=\bigoplus_Y\left(\cI_{\mathcal U^{(Y)}(\mathrm{SU}(d))}\otimes\cD_{\mathcal U^{(Y)}(S_{N+M})}\right)[\Pi_Y\,\rho],
\label{supp:twilringnice}
\end{align}
where $\cI$ is the identity map, i.e., $\cI_{\cH}[A]=A$, $\cD$ is the completely depolarizing map, i.e., $\cD_{\cH}[A]=\frac{\tr A}{\mathrm{dim}(\cH)}
\one_{\cH}$, and $\Pi_Y$ is the projector onto the block $\mathcal U^{(Y)}(\mathrm{SU}(d))\otimes \mathcal U^{(Y)}(S_{N+M})$.  

In the following we will need some more observations. Define $\mathrm{Sym}_{N}^{(d)}$ as the completely symmetric subspace of $\mathcal H_d^{\otimes N}$; in particular, if taken $Y_{max}$ to be the label for a Young diagram of just one row, one has $\mathrm{Sym}_{N}^{(d)}=\mathcal U^{(Y_{max})}(\mathrm{SU}(d))\otimes\mathcal U^{(Y_{max})}(S_{N})$. Moreover, the tensor product space 
\be\label{eq:smn}
\mathrm{Sym}_{N,M}^{(d)}:=\mathrm{Sym}_{N}^{(d)}\otimes\mathrm{Sym}_{M}^{(d)}\subset\cH_{d}^{\otimes{N+M}}.
\ee

admits a decomposition 
\be\label{decompositionsym}
\mathrm{Sym}_{N,M}^{(d)}=\oplus_{J}\left(\mathcal U^{(J)}(\mathrm{SU}(d))\otimes \mathcal K^J\right),
\ee

where $J$ are the labels of Young diagram with two rows and $\mathcal K^J:=\mathrm{span}\{ \ket{J}\}$, $\ket{J}\in \mathcal U^{(J)}(S_{M+N})$, $\mathcal K^J$ being a one dimensional space because the multiplicity of each irrep $\mathcal U^{(J)}(\mathrm{SU}(d))$ is one. Notice that the label $J$ can be indexed by an half-integer $\frac{M-N}{2}\leq J \leq\frac{M+N}{2}$.


The states of the form $\left(\proj{\psi}\right)^{\otimes N}\otimes\left(\proj{\phi}\right)^{\otimes M}$  are supported on $\mathrm{Sym}_{N,M}^{(d)}$. Define also $\mathcal E:=\mathrm{span}\{\ket{\psi},\ket{\psi_\perp}\}\subseteq\cH_d$. Since $\mathcal E$ has dimension 2, it is isomorphic to $\mathcal H_2$. Using this isomorphism we can define $\mathrm{Sym}^{\mathcal E}_N\cong \mathrm{Sym}^{(2)}_N$. Notice that $\left(\proj{\psi}\right)^{\otimes N}\otimes\left(\proj{\phi}\right)^{\otimes M}$ is also supported on the product of the two completely symmetric subspaces of $\cE^{\otimes N}$ and $\cE^{\otimes M}$, $\mathrm{Sym}^{\mathcal E}_{N,M}:=\mathrm{Sym}^{\mathcal E}_{N}\otimes \mathrm{Sym}^{\mathcal E}_M$, which is isomorphic to
\be\label{decompositionsym20}
\mathrm{Sym}_{N,M}^{(2)}:=\oplus_{J}\left(\mathcal U^{(J)}(\mathrm{SU}(2))\otimes \mathcal K^{(2)}_{J}\right)\subseteq\mathrm{Sym}_{N,M}^{(d)}.
\ee 
where now $K^{(2)}_{J}=\mathrm{span}\{ \ket{J}_{(2)}\}$, $\ket{J}_{(2)}\in \mathcal U^{(J)}(S_{M+N})$. Using this isomorphism, we can write the decomposition

\be\label{decompositionsym2}
\mathrm{Sym}_{N,M}^{\mathcal E}:=\oplus_{J}\left(\mathcal U^{(J)}_{\mathcal E}(\mathrm{SU}(2))\otimes \mathcal K_{\mathcal E}^{J}\right)\subseteq\mathrm{Sym}_{N,M}^{(d)}.
\ee 
and now $K_{\mathcal E}^{J}=\mathrm{span}\{ \ket{J}_{\mathcal E}\}$, $\ket{J}_{\mathcal E}\in \mathcal U^{(J)}(S_{M+N})$.
What is more, $\mathcal E^{\otimes N}\otimes \mathcal E ^{\otimes M}$ is an invariant subspace of $\cH_{d}^{\otimes M+N}$ under the action of the symmetric group, therefore the irreps of the symmetric group supported on $\mathcal E ^{\otimes N}\otimes \mathcal E ^{\otimes M}$ can also be taken as irreps in the decomposition of the representation of $S_{M+N}$ on $\cH_{d}^{\otimes M+N}$, and it follows that $\mathcal U^{(J)}_{\mathcal E}(\mathrm{SU}(2))\otimes  \mathcal K_{\mathcal E}^{J} \subseteq \mathcal U^{(J)}(\mathrm{SU}(d))\otimes\mathcal U^{(J)}(\mathcal S_{N+M})$. 
Finally, since $\mathrm{Sym}_{N,M}^{(2)}\subseteq\mathrm{Sym}_{N,M}^{(d)}$ and each of the $\mathcal U^{(J)}_{\mathcal E}(\mathrm{SU}(2))\otimes \mathcal K^{J}_{\mathcal E}$ and $\mathcal U^{(J)}(\mathrm{SU}(d))\otimes \mathcal K^{J}$ are included in the same subspace $\mathcal U^{(J)}(\mathrm{SU}(d))\otimes\mathcal U^{(J)}(\mathcal S_{N+M})$, it follows that
\be\label{2dincl}
\mathcal U^{(J)}_{\mathcal E}(\mathrm{SU}(2))\otimes \mathcal K_{\mathcal E}^{J}\subseteq\mathcal U^{(J)}(\mathrm{SU}(d))\otimes \mathcal K^{J}
\ee
and $\ket{J}=\ket{J}_{\mathcal E}$ for each $J$.

\section{Average state at fixed overlap}
\label{app:overlap_pure_states}
In this section we derive the form of the average state of $N$ copies of a pure state $\ket{\psi}\in\cH_d$ and 
$M\geq N$ copies of a pure state $\ket{\phi}\in\cH_d$ with $\ket{\psi}$ and $\ket{\phi}$ having fixed overlap. Using group representation theory techniques we find a basis in which the average state is diagonal and compute the eigenvalues. We generalize the result to the average state in the unlabeled scenario.

Without loss of generality we may write the latter of the two states as 
\be
\ket{\phi}=V(\theta)\ket{\psi}\equiv\sqrt{c}\ket{\psi}+\sqrt{1-c}\ket{\psi_\perp}
\label{app:dto2iso}
\ee
where $c=\cos^2\frac{\theta}{2}\in(0,1),\, \theta\in(0,\pi), \,
\braket{\psi_\perp}{\psi}=0$. 

As both $\ket{\psi},\,\ket{\phi}\in\cH_d$ are randomly chosen the global state describing the $N+M$ qudits is given by
\be
\cG_{\mathrm{SU}(d)}\left[\left(\proj{\psi}\right)^{\otimes N}\otimes\left(\proj{\phi}\right)^{\otimes M}\right]\equiv\int_{\mathrm{SU}(d)} \d U\, \left(U\proj{\psi}U^\dagger\right)^{\otimes N}\otimes \left(U\proj{\phi}U^\dagger\right)^{\otimes M}
\label{app:su(d)twirling}
\ee
where $dU$ is the Haar measure of $SU(d)$. We can decompose the integral over $\mathrm{SU}(d)$ in Eq.~\eqref{app:su(d)twirling} as 
follows. First we consider the two-dimensional subspace $\mathcal E\equiv\mathrm{span}\{\ket{\psi},\ket{\psi_\perp}\}\subseteq\cH_d$. Using the invariance of the Haar measure, we first perform the group average over the $\mathrm{SU}(2)$ subgroup of $\mathrm{SU}(d)$ which acts non-trivially only on $\mathcal E$, which we denote as $\mathrm {SU}(\mathcal E)$.  Afterwards, we can average over all $\mathrm{SU}(d)$. This implies that Eq.~\eqref{app:su(d)twirling} can be 
written as 
\begin{align}\nonumber
&\cG_{\mathrm{SU}(d)}\left[\left(\proj{\psi}\right)^{\otimes N}\otimes\left(\proj{\phi}\right)^{\otimes M}\right]
=\int_{\mathrm{SU}(d)} \d U\,U^{\otimes(N+M)}\left[\left(\proj{\psi}\right)^{\otimes N}\otimes\left(\proj{\phi}\right)^{\otimes M}\right]\, {U^\dagger}^{\otimes(N+M)}\\\nonumber
&=\int_{\mathrm{SU}(d)}\d U\, U^{\otimes (N+M)}\left(\int_{\mathrm{SU}(\mathcal E)}\d V\, V^{\otimes(N+M)}\left[\left(\proj{\psi}\right)^{\otimes N}\otimes\left(\proj{\phi}\right)^{\otimes M}\right]{V^\dagger}^{\otimes(N+M)}\right){U^\dagger}^{\otimes (N+M)}\\
&=\cG_{\mathrm{SU}(d)}\left[\cG_{\mathrm{SU}(\mathcal E)}\left[\left(\proj{\psi}\right)^{\otimes N}\otimes\left(\proj{\phi}\right)^{\otimes M}\right]\right]
\label{app:compositionoftwirls}
\end{align}

First of all we compute the action of $\cG_{\mathrm{SU}(2)}$ on $\left(\proj{\psi}\right)^{\otimes N}\otimes\left(\proj{\phi}\right)^{\otimes M}$, when $d=2$. We use the addition rules for angular momentum on $\cH_{2}^{\otimes N+M}$ to write 
\begin{align}\nonumber
\ket{\psi}^{\otimes N}\otimes\ket{\phi}^{\otimes M}&=\ket{\frac{N}{2},\frac{N}{2}}\otimes \sum_{k=-\frac{M}{2}}^{\frac{M}{2}}d^{\left(\frac{M}{2}\right)}_{k\frac{M}{2}}\ket{\frac{M}{2},k}\\
&=\sum_{J=J_{\min}}^{J_{\max}}\sum_{k=-\frac{M}{2}}^{\frac{M}{2}}C^{J,\frac{N}{2}+k}_{\frac{N}{2},\frac{N}{2};\frac{M}{2},k}
d^{\left(\frac{M}{2}\right)}_{k,\frac{M}{2}}(\theta)\,\ket{J,\frac{N}{2}+k},
\label{app:totalJstate}
\end{align}
where $C^{J,\frac{N}{2}+k}_{\frac{N}{2},\frac{N}{2};\frac{M}{2},k}=\braket{J, \frac{N}{2}+k}{\frac{N}{2},\frac{N}
{2};\frac{M}{2},k}$ are the Clebsch-Gordan coefficients. Using Eq.~\eqref{supp:su(d)twirl} and the decomposition in Eq. \eqref{decompositionsym2} ($\left(\proj{\psi}\right)^{\otimes N}\otimes\left(\proj{\phi}\right)^{\otimes M}$ is supported on $\mathrm{Sym}_{N,M}^{\mathcal E}$) the first Haar-measure group average of Eq.~\eqref{app:compositionoftwirls} reads
\be
\cG_{\mathrm{SU}(2)}\left[\left(\proj{\psi}\right)^{\otimes N}\otimes\left(\proj{\phi}\right)^{\otimes M}\right]=\sum_{J=J_{\min}}^{J_{\max}}p(J|c)\frac{\one_{\mathcal U^{(J)}(\mathrm{SU}(2))}}{2J+1}\otimes \proj{J}_{\mathcal U^{(J)}(S_{N+M})},
\label{app:overlap_pure}
\ee
where we have used the total angular momentum $J$, instead of the Young frame label $Y$, $\chi_J=2J+1$, and $\proj{J}_{\mathcal U^{(J)}(S_{N+M})}$ is a pure state.
The coefficients of Eq.~\eqref{app:overlap_pure} are:
\begin{align}\nonumber
p(J|c)&=\sum_{k=-\frac{M}{2}}^{\frac{M}{2}}\left(C^{J,\frac{N}{2}+k}_{\frac{N}{2},\frac{N}{2};\frac{M}{2},k}\, D^{\left(\frac{M}{2}\right)}_{k\frac{M}{2}}(2\arccos\sqrt{c})\right)^2\\   \nonumber
&=\frac{(2J+1)(J+J_{\min})!N!}{(J-J_{\min})!(J_{\max}-J)!(J_{\max}+1+J)!}\sum_{k=-\frac{M}{2}}^{J-\frac{N}{2}}\frac{(\frac{M}{2}-k)!(J+\frac{N}{2}+k)!}{(J-\frac{N}{2}-k)!(\frac{M}{2}+k)!} \left(D^{\left(\frac{M}{2}\right)}_{k\frac{M}{2}}(2\arccos\sqrt{c})\right)^2\\  \nonumber
&=\frac{(2J+1)(J+J_{\min})!N!M!}{(J-J_{\min})!(J_{\max}-J)!(J_{\max}+1+J)!}\sum_{k=-\frac{M}{2}}^{J-\frac{N}{2}}\frac{(J+\frac{N}{2}+k)!}{(J-\frac{N}{2}-k)!(\frac{M}{2}+k)!^2}(1-c)^{\frac{M}{2}-k}c^{\frac{M}{2}+k}\\
&=\frac{(2J+1)N!M!(1-c)^{M}}{(J_{\max}-J)!(J_{\max}+1+J)!}P_{J+J_{\min}}^{(0,-2 J_{\min})}\left(\frac{1+c}{1-c}\right),
\label{app:probJgivenc}
\end{align}
and we have made use of the following expression for the Wigner $D$ matrix in going from the second to the third line in 
Eq.~\eqref{app:probJgivenc}
\begin{align}
D^{\left(J\right)}_{z',z}(\theta)&=\sqrt{\frac{(J+z)!(J-z)!}{(J+z')!(J-z')!}}\sin^{(z-z')}\left(\frac{\theta}{2}\right)\cos^{(z+z')}
\left(\frac{\theta}{2}\right)\, P_{(J-z)}^{(z-z',z+z')}(\cos\theta),
\label{app:Wignerd}
\end{align}
with $P_n^{(\alpha,\beta)}(x)$ the Jacobi polynomials, defined in general as
\be
P_{n}^{(\alpha, \beta)}(x)=\frac{\Gamma(\alpha+n+1)}{n ! \Gamma(\alpha+\beta+n+1)} \sum_{m=0}^{n}\left(\begin{array}{c}{n} \\ {m}\end{array}\right) 
\frac{\Gamma(\alpha+\beta+n+m+1)}{\Gamma(\alpha+m+1)}\left(\frac{x-1}{2}\right)^{m}.
\ee

Let's consider now $d>2$. Notice that $\mathcal E^{\otimes M+N}\cong \oplus_J \mathcal U^{(J)}_{\mathcal E}(\mathrm{SU}(2))\otimes \mathcal U_{\mathcal E}^{(J)}(S_{N+M})$. Performing the average over $\mathrm{SU}(d)$ on the state given by Eq.~\eqref{app:overlap_pure} results in 
\begin{align}\nonumber
\cG_{\mathrm{SU}(d)}\left[\cG_{\mathrm{SU}(\mathcal E)}\left[\left(\proj{\psi}\right)^{\otimes N}\otimes\left(\proj{\phi}\right)^{\otimes M}\right]\right]&=\bigoplus_{Y}\, \left(\cD_{\mathcal U^{(Y)}(\mathrm{SU}(d))}\otimes \cI_{\mathcal U^{(Y)}(S_{N+M})}\right) \left[\Pi_Y\left(\sum_{J=J_{\min}}^{J_{\max}}p(J|c)\frac{\one_{\mathcal U^{(J)}_{\mathcal E}(\mathrm{SU}(2))}}{2J+1}\right.\right.\\
&\left.\left.\otimes\proj{J}_{\mathcal U^{(J)}_{\mathcal E}(S_{N+M})}\right)\right]\nonumber\\
&= \sum_{J=J_{\min}}^{J_{\max}}p(J|c)\frac{\one_{\mathcal U^{(J)}(\mathrm{SU}(d))}}{\chi_{J}}\otimes\proj{J}_{\mathcal U^{(J)}(S_{N+M})},
\label{eq:average_state_diag}
\end{align}
where in the last equality we used the fact that the support of $\one_{\mathcal U^{(J)}_{\mathcal E}(\mathrm{SU}(2))}\otimes \proj{J}_{\mathcal U^{(J)}_{\mathcal E}(S_{N+M})}$ is a subspace of the support of $\one_{\mathcal U^{(J)}(\mathrm{SU}(d))}\otimes\proj{J}_{\mathcal U^{(J)}(S_{N+M})}$, as stated in Eq. \ref{2dincl}. 

Hitherto in the computation we tacitly assumed that the copies of states are labelled, i.e., that the first $N$ states are all $\ket{\psi}\in\cH_d$ and the remaining $M$ 
states are all $\ket{\phi}\in\cH_d$.  We now lift this assumption and derive the form of the average state in the case where the copies of the states are unlabelled.
This is equivalent to averaging over all possible permutations of the $N+M$ copies, i.e., by applying the map $\cG_{S_{N+M}}[\rho(c)]$ of 
Eq.~\eqref{supp:twilringnice}.  The final state can be easily shown to be 
\begin{align}\nonumber
\rho(c)_{ul}&:=\cG_{S_{N+M}}[\rho(c)]=\cG_{S_{N+M}}\circ\cG_{\mathrm{SU}(d)}\left[\left(\proj{\psi}\right)^{\otimes N}\otimes\left(\proj{\phi}\right)^{\otimes M}\right]\\
&=\sum_{J=J_{\min}}^{J_{\max}} p(J|c) \frac{\one_{\mathcal U^{(J)}(\mathrm{SU}(d))}}{\chi_J}\otimes \frac{\one_{\mathcal U^{(J)}(S_{M+N})}}{\omega_J},
\label{supp:unlabelled}
\end{align}
by observing that the maps $\cG_{\mathrm{SU}(d)},\, \cG_{S_{N+M}}$ both project on the same irrep label $J$, but depolarize 
the states on different irrep subspaces 
$\mathcal U^{(J)}(\mathrm{SU}(d))$ and $\mathcal U^{(J)}(S_{N+M})$ respectively.  As all the information concerning the overlap between the two states is extracted by the projective measurement $\{\Pi_J\}_J$, and the action of any permutation--or global unitary rotation--does not alter the statistics $p(J|c)$. Consequently, all our results apply equally well to both labelled and unlabelled scenarios.


\section{Fisher Information of $p(J|c)$}\label{FIasy}
In this section we derive the Fisher information of the probability distribution given by Eq.~\eqref{app:probJgivenc}, and analyze its asymptotics. We can obtain the first non trivial order by using an asymptotic approximation for the Jacobi polynomials and large deviation evaluations. Using the same techniques, we identify an asymptotically unbiased estimator that saturates the Cramer-Rao bound.

The result we obtain is summarized by the following theorem:

\begin{theorem*}
Given $N$ and $M$ copies of two Haar-random states, respectively $\ket{\psi}$ and $\ket{\phi}$, define $\ket{\Psi}=\ket{\psi}^{\otimes N}\otimes\ket{\phi}^{\otimes M}$. The mean square error on the estimation of the overlap $c=|\braket{\psi}{\phi}|^2$ attained by the weak Schur sampling projectors $\{E_J\}_J$ with the estimator $c_J=\left(\frac{J}{M+N}\right)^2$ is
\be
v(c)=\sum_{J}\tr{E_J\proj{\Psi}}(c_J-c)^2=\frac{4c(1-c)}{N+M}+\mathcal{O}\left((N+M)^{-\frac{3}{2}}\right)
\ee

This measurement is optimal in the sense that saturates the quantum Cramer-Rao bound for the problem of estimating of $c$ from the family of average states at fixed overlap

\be
\rho(c)=\int\d U \, U^{\otimes(N+M)}\, \proj{\Psi}\, U^{\dagger \otimes(N+M)}.
\ee

This implies that $c$ can be estimated at a relative error $\frac{\sqrt{v(c)}}{c}=\epsilon$ with $N=M=\mathcal O(\frac{1}{\epsilon^2 c})$ samples.

\end{theorem*}

We begin by recalling the definition of the Fisher information 
\be
H(c)=\sum_{J=J_{\min}}^{J_{\max}}p(J|c)\left(\frac{\frac{\d p(J|c)}{\d c}}{p(J|c)}\right)^2=\sum_{J=J_{\min}}^{J_{\max}}\frac{\left(\frac{\d p(J|c)}{\d c}\right)^2}{p(J|c)}.
\label{app:FI}
\ee
Using the identity
\be
\frac{\d^m P_n^{(\alpha,\beta)}(x)}{\d x^m}=\frac{(\alpha+\beta+n+m)!}{2^m(\alpha+\beta+n)!}P_{n-m}^{(\alpha+m,\beta+m)}(x),
\label{app:Jacobiderivatives}
\ee
it follows that 
\begin{align}
\frac{\d p(J|c)}{\d c}&=\frac{(2 J+1) M! N! (1-c)^{M-2}}{\left(J_{\max}-J\right)!
   \left(J+J_{\max}+1\right)!}\nonumber\\
   &\times \left((J-J_{\min}+1)
   P_{J+J_{\min}-1}^{(1,1-2J_{\min})}\left(\frac{1+c}{1-c}\right)-(1-c) M
   P_{J+J_{\min}}^{(0,-2J_{\min})}\left(\frac{1+c}{1-c}\right)\right).
\label{app:derivPj}
\end{align}
For $x \geq 1$ the following asymptotic expansion for the Jacobi polynomials holds~\cite{Szego1939,Darboux1878}. Defining the function
\be
Q_{n}^{(\alpha,\beta)}(x)=\frac{\left(\sqrt{x+1}+\sqrt{x-1}\right)^{\alpha+\beta}\left(x+\sqrt{x^2-1}\right)^{n+\frac{1}{2}}}{\sqrt{2\pi n}\,(\sqrt{x-1})^{\alpha}\,(\sqrt{x+1})^{\beta}\,\sqrt[4]{x^2-1}}
\ee 
one has
\be
P_{n}^{(\alpha,\beta)}(x)=Q_{n}^{(\alpha,\beta)}(x)\left(1+\mathcal O\left(\frac 1 n\right)\right),
\label{app:Jacobiasymptotic}
\ee
where the convergence is uniform on any half-line $x\in [1+\delta, +\infty)$, $\delta>0$.
We want to use the leading order as an approximation to the Jacobi polynomial. Notice that the remainder in the expansion does not depend on $J_{\max}$. Take $J_0$ such that for each $J> J_0$ and a certain $C>0$
\be
\left|\frac{P_{J+J_{\min}}^{(0,-2J_{\min})}(x)}{Q_{J+J_{\min}}^{(0,-2J_{\min})}(x)}-1 \right |+\left|\frac{P_{J+J_{\min}-1}^{(1,1-2J_{\min})}(x)}{Q_{J+J_{\min}-1}^{(1,1-2J_{\min})}(x)}-1\right |\leq \frac {C}{J+J_{\min}}.
\ee

We can then split the sum in Eq. \ref{app:FI} in two parts. The first is
\be
H(c)_1:=\sum_{J=J_{\min}}^{J_{0}}p(J|c)\left(\frac{\frac{\d p(J|c)}{\d c}}{p(J|c)}\right)^2.
\ee
Writting $p(J|c)$ as
\be
p(J|c)=\frac{2J+1}{2J_{\max}+1}\frac{\binom{2J_{\max}+1}{J_{\max}-J}}{\binom{2J_{\max}}{J_{\max}-J_{\min}}}(1-c)^{J_{\max}-J} \sum_{s=0}^{J-J_{\min}}\binom{J+J_{\min}}{s}\binom{J-J_{\min}}{s}c^s.
\ee
and noting that
\be
\left(\frac{\frac{\d p(J|c)}{\d c}}{p(J|c)}\right)=-\frac{(J_{\max}-J)}{1-c}+ \frac{(1-c)^{J_{\max}-J}\sum_{s=0}^{J-J_{\min}}s\binom{J+J_{\min}}{s}\binom{J-J_{\min}}{s}c^{s}}{ c p(J|c)}\leq -\frac{(J_{\max}-J)}{1-c}+ \frac{J-J_{\min}}{c},
\ee
it follows that for $J\leq J_0$
\be
\left|\left(\frac{\frac{\d p(J|c)}{\d c}}{p(J|c)}\right)\right|\leq \frac{c J_{\max}+J_0+ (1-c)J_{\min}}{c(1-c)}.
\ee

On the other hand, isolating in $p(J|c)$ the terms depending on $J_{\max}$, and defining $w(J,J_{\min},c)=(2J+1)\sum_{s=0}^{J-J_{\min}}\binom{J+J_{\min}}{s}\binom{J-J_{\min}}{s}c^s$ which is bounded for $J_{\min}\leq J \leq J_0$, we have
\be
p(J|c)=\frac{1}{2J_{\max}+1}\frac{\binom{2J_{\max}+1}{J_{\max}-J}}{\binom{2J_{\max}}{J_{\max}-J_{\min}}}(1-c)^{J_{\max}-J}w(J,J_{\min},c)\\
\label{supp:JstatJmax}
\ee
By introducing the following binomial distribution
\be
q(J):=\mathrm{Bin}(2J_{\max}+1,J_{\max}-J, p)=\binom{2J_{\max}+1}{J_{\max}-J}p^{J_{\max}-J}(1-p)^{J_{\max}+J+1},
\label{app:binomial}
\ee
with $p=\frac{1-\sqrt{c}}{2}$, whose mean and variance are given by 
\begin{align}\nonumber
\mu&=(2J_{\max}+1)\left(\frac{1-\sqrt{c}}{2}\right)\\
\sigma^2&=\frac{(2J_{\max}+1)}{4}(1-c).
\label{app:binommeanvar}
\end{align}
Eq.~\eqref{supp:JstatJmax} can be written as 
\be
p(J|c)=\frac{1}{2J_{\max}+1}\frac{2^{2J_{\max}}}{\binom{2J_{\max}}{J_{\max}-J_{\min}}}q(J) w'(J,J_{\min},c),
\ee
with $w'(J,J_{\min},c)$ still bounded.  

Now by Sanov's theorem~\cite{Sanov1957}, the binomial distribution  $q(J)$ is exponentially suppressed as $J_{\max}\rightarrow \infty$ 
and $J<J_0$. In particular, if $J_{\max}>\frac 1{\sqrt c}J_{0}$
\be\label{sanov}
q(J)\leq \exp^{-(2J_{\max})D(\frac{J_{\max}-J_0}{2J_{\max+1}}||\frac{1-\sqrt{c}}{2})},
\ee
where $D(\frac{J_{\max}-J_0}{2J_{\max}+1}||\frac{1-\sqrt{c}}{2})$ is the relative entropy between the two Bernoulli distributions with 
probabilities $p_1=\frac{J_{\max}-J_0}{2J_{\max+1}}$ and $p_2=\frac{1-\sqrt{c}}{2}$. For $J_{\max}\rightarrow \infty$, $p_1\rightarrow 
\frac 1 2,\, D(\frac{J_{\max}-J_0}{2J_{\max}+1}||\frac{1-\sqrt{c}}{2})\rightarrow D(\frac 1 2 ||\frac{1-\sqrt{c}}{2})>0$ unless $c=0$. 
As the rest of the terms in both $p(J|c)$ and $\left|\left(\frac{\frac{\d p(J|c)}{\d c}}{p(J|c)}\right)\right|$ tend to a power law in $J_{max}$, 
$H(c)_1$ is exponentially suppressed. 

As an indication for which ratio $ \frac{J_{\max}}{J_{\min}}$ is expected to give exponential suppression when $c$ is small, we can 
look at the first order of the Taylor expansion of $D(p_1||p_2)$ in $\frac 1 {J_{\max}}$ and $\sqrt{c}$ around zero: $D(p_1||p_2)=-\frac 1 2 \log(1-c)-\frac 1 4 \frac{2J_0+1}{J_{max}} \log\frac {1+
\sqrt{c}}{1-\sqrt{c}}$, which for small $c$ requires $J_{\max}c>>1$ from the zeroth order and $J_{\max}>>\frac{1}{\sqrt{c}}J_0$ from 
the first order, implying also $J_{\max}>>\frac{1}{\sqrt{c}}J_{\min}$. 

It remains to evaluate 
\be
H(c)_2:=\sum_{J=J_0+1}^{J_{\max}}p(J|c)\left(\frac{\frac{\d p(J|c)}{\d c}}{p(J|c)}\right)^2.
\label{supp:Hcpart2}
\ee
Using Eq.~\eqref{app:Jacobiasymptotic} one obtains, after some algebra
\be\begin{aligned}
\frac{\left(\frac{\d p(J|c)}{\d c}\right)^2}{p(J|c)}&=\frac{(2 J+1) M! N! \left(1+\sqrt{c}\right)^{2 J-1}(1-c)^{J_{\max}-J}}{2 \sqrt{\pi }\left(1-\sqrt{c}\right)^2 c^{5/4} (J+J_{\min}-1) \sqrt{J+J_{\min}} \left(J_{\max}-J\right)!\left(J+J_{\max}+1\right)!}\\
&\times\left((J-J_{\min}+1)\sqrt{J+J_{\min}}-M \sqrt{c (J+J_{\min}-1)}\right)^2\left(1+\mathcal O\left(\frac 1 {J+J_{\min}}\right)\right),
\label{app:asymptoticFisher}
\end{aligned}\ee
where the same binomial distribution, $q(J)$ of Eq.~\eqref{app:binomial}, appears again. Writing 
\be
\frac{\left(\frac{\d p(J|c)}{\d c}\right)^2}{p(J|c)}=f(M,N,c) g(J,J_{\max},J_{\min},c)q(J)\left(1+\mathcal O\left(\frac 1 {J+J_{\min}}\right)\right)
\label{app:fisher2}
\ee
where
\begin{align}\nonumber
f(M,N,c)&=\frac{M! N! 2^{2J_{\max}}}{\sqrt{\pi } (1-c)^2 c^{5/4}(2J_{\max}+1)!}\\
g(J,J_{\max},J_{\min},c)&=\frac{(2 J+1) \left((J-J_{\min}+1)\sqrt{J+J_{\min}}-M \sqrt{c (J+J_{\min}-1)}\right)^2}{(2J_{\max}+1)(J+J_{\min}-1) \sqrt{J+J_{\min}}}\nonumber\\
&=\frac{(2 J+1)(J-J_{\min}+1)}{(2J_{\max}+1)(J+J_{\min}+1)}\sqrt{J+J_{\min}}+ c\frac{(2 J+1)(J_{\max}+J_{\min})^2}{(2J_{\max}+1)\sqrt{J+J_{\min}}}\nonumber \\&-\sqrt c\frac{(2 J+1)(J-J_{\min}+1)}{(2J_{\max}+1)\sqrt{J+J_{\min}-1}}
\end{align}
we have
\be
H(c)_2=f(M,N,c)\sum_{J=J_0+1}^{J_{\max}}g(J,M,N,c)\mathrm{Bin}(2J_{\max}+1,J_{\max}-J, p)\left(1+\mathcal O\left(\frac 1 {J+J_{\min}}\right)\right).
\label{app:explicitQFI}
\ee
In the limit $J_{\max}\to\infty$
\be
f(M,N,c)=\frac{1}{2(1-c)^2 c^{\frac{5}{4}}}\frac{1}{\sqrt{J_{\max}}}+\mathcal{O}\left(J^{-\frac{3}{2}}_{\max}\right).
\label{app:fseries}
\ee
whilst $g(J, J_{\max}, J_{\min},c)=\mathcal O (J_{\max}^{\frac 3 2})$ by inspection. 

Now consider the first $\sqrt{d}J_{\max}-(J_0+1)$ terms in Eq.~\eqref{supp:Hcpart2}, where $d<c$, 
\be
H(c)_{2a}:=\sum_{J=J_0+1}^{\sqrt{d} J_{\max}-1}p(J|c)\left(\frac{\frac{\d p(J|c)}{\d c}}{p(J|c)}\right)^2.
\ee
Sanov's theorem tells us that these are exponentially suppressed, 
\be
H(c)_{2a}\leq \exp^{-(2J_{\max})D(\frac{1-\sqrt{d}}{2}||\frac{1-\sqrt{c}}{2})} \mathcal O (J_{\max}^2),
\ee
and as a result we are only left with having to evaluate
\be
H(c)_{2b}:=\sum_{J=\sqrt{d}J_{\max}}^{J_{\max}}q(J) g(J,J_{\max},J_{\min},c)\left(1+\mathcal O\left(\frac 1 {J+J_{\min}}\right)\right).
\label{supp:Hcpart2b}
\ee
As we are interested in computing the $QFI$ in the limit $J_{\max}\to\infty$ we will perform a Taylor series 
expansion of $H(c)_{2b}$, around the mean value $\langle J\rangle=J_{\max}-\mu=\sqrt{c}\left(J_{\max}+\frac{1}{2}\right)-\frac{1}{2}
$ of $q(J)$. For any analytical function $h(J)$, using the Lagrange remainder, an expansion about $\langle J\rangle$ is given by 
\be
\sum_{J=\sqrt{d}J_{\max}}^{J_{\max}}q(J)h(J)=\sum_{J=\sqrt{d}J_{\max}}^{J_{\max}}q(J)\left(\sum_{s=0}^2\left. \frac{(J-\langle J\rangle)^s}{s!}\frac{\d^s h(J)}{\d J^s}\right\rvert_{J=\langle J\rangle}+\left. \frac{(J-\langle J\rangle)^3}{3!}\frac{\d^3 h(J')}{\d J'^3}\right\rvert_{J'=\xi(J)}\right),
\label{app:Tayloraboutmean}
\ee
with $J\leq \xi(J)\leq \langle J\rangle$ (so that in any case $\sqrt{d}J_{\max}\leq\xi(J)\leq J_{\max}$).
Using Eq.~\eqref{app:Tayloraboutmean} and the fact that $g(J,J_{\max},J_{\min},c)$ is analytic for $J>J_{\min}$ by inspection, we have
\begin{multline}\label{taylorlagrange}
\sum_{J=\sqrt{d}J_{\max}}^{J_{\max}}q(J)g(J,J_{\max},J_{\min},c)=\sum_{s=0}^2 \frac{1}{s!}\left.\frac{\d^s g(J,J_{\max},J_{\min},c)}{\d J^s}\right\rvert_{J=\langle J\rangle}\left(\sum_{J=\sqrt{d}J_{\max}}^{J_{\max}}q(J) \left(J-\langle J\rangle\right)^s\right)\\+ \frac{1}{3!}\left.\frac{\d^3 g(J',J_{\max},J_{\min},c)}{\d J'^3}\right\rvert_{J'=\xi(J)}\left(\sum_{J=\sqrt{d}J_{\max}}^{J_{\max}}q(J) \left(J-\langle J\rangle\right)^s\right),
\end{multline}
where the change in the order of the summation is allowed since one of the sums is finite. Applying Sanov's theorem, Eq.~\eqref{sanov}, gives
\begin{multline}
\sum_{J=-J_{\max}-1}^{\sqrt{d}J_{\max}}q(J)(J-\langle J\rangle)^s\leq (2J_{\max}+1) (J_{\max}+\langle J\rangle+1)^s\exp^{-(2J_{\max})D(\frac{J_{\max}-\sqrt{d}J_{\max}}{2J_{\max+1}}||\frac{1-\sqrt{c}}{2})}\\=(2J_{\max}+1) \left(J_{\max}-\langle J\rangle\right)^s\left(\frac{J_{\max}+\langle J\rangle+1}{J_{\max}-\langle J\rangle}\right)^s  \exp^{-(2J_{\max})D(\frac{J_{\max}-\sqrt{d}J_{\max}}{2J_{\max+1}}||\frac{1-\sqrt{c}}{2})}\\=(2J_{\max}+1) \left(J_{\max}-\langle J\rangle\right)^s\left(\frac{1+\sqrt{c}}{1-\sqrt{c}}\right)^s\exp^{-(2J_{\max})D(\frac{J_{\max}-\sqrt{d}J_{\max}}{2J_{\max+1}}||\frac{1-\sqrt{c}}{2})}.
\end{multline}
It follows that we can change the summation in $J$ of Eq.~\eqref{taylorlagrange} to start from $-J_{\max}-1$ just by adding a term 
exponentially suppressed in $J_{\max}$. For the Lagrange remainder, before extending the summation, we notice that for $
\sqrt{d}J_{\max}\leq\xi(J)\leq J_{\max}$, $\left.\frac{d^s g(J',M,N,c)}{d J'^s}\right\rvert_{J'=\xi(J)}= \mathcal O(J_{\max}^{\frac 3 2- n})$ 
(by inspection).

Finally, using Eq.~\eqref{app:binommeanvar} and some algebra one obtains
\be\begin{aligned}
\sum_J g(J,J_{\max},J_{\min},c)q(J)&=g(\langle J\rangle ,J_{\max},J_{\min},c)+\frac{1}{2}\sigma^2\left.\frac{\d^2 g(J,J_{\max},J_{\min},c)}{\d J^2}\right\rvert_{J=\langle J\rangle}+\mathcal{O}(J^{-\frac{1}{2}}_{\max})\\
&=\frac{1}{2} (1-c) c^{\frac 14} J_{\max}^{\frac 32}+\mathcal{O}(J^{\frac{1}{2}}_{\max}).
\label{app:gseries}
\end{aligned}\ee
Multiplying Eqs.~(\eqref{app:fseries},~\eqref{app:gseries}) gives the final result
\be
H(c)=\frac{J_{\max}}{2c(1-c)}+\mathcal{O}(1)=\frac{M+N}{4c(1-c)}+\mathcal{O}(1).
\label{app:FI_final}
\ee
The neglected term $\mathcal O \left(\frac{1}{J+J_{\min}}\right)$ in Eq. (\ref{app:explicitQFI}) contributes to $H(c)$ with a next to leading order term, by simple power counting, adapting the steps above.

Using similar techniques it can be shown that $H(c)$ is achievable by the estimator of Eq.~(7) of the main text.  First note that the estimator is indeed asymptotically unbiased since 
\begin{align}\nonumber
\langle c_{\rm op}^{\rm loc}\rangle&:=\sum_{J=J_{\min}}^{J_{\max}}\left(\frac{J}{J_{\max}}\right)^2p(J|c)\\
&=\frac{2^{2J_{\max}}M!N!}{c^{\frac{1}{4}}\sqrt{\pi}(2J_{\max}+1)!J^2_{\max}}\sum_{J=\sqrt{d}J_{\max}}^{J_{\max}}\frac{J^2(2 J+1)}{\sqrt{J-J_{\min}}}\mathrm{Bin}(2J_{\max}+1,J_{\max}-J, p),
\label{app:asymptoticestimator}
\end{align}
apart from an exponentially suppressed term, and we have again used the asymptotic expansion of Eq.~\eqref{app:Jacobiasymptotic} and identified the binomial distribution of Eq.~\eqref{app:binomial}.
In the limit $J_{\max}\to\infty$ one can proceed as above, separating the $J$-dependence and applying Eq.~\eqref{app:Tayloraboutmean}, to obtain
\begin{align}\nonumber
&\frac{2^{2J_{\max}}M!N!}{c^{\frac{1}{4}}\sqrt{\pi}(2J_{\max}+1)!J_{\max}^2}=\frac{1}{2c^{\frac{1}{4}}}\frac{1}{J^{\frac{5}{2}}_{\max}}+\mathcal{O}(J^{-\frac 72}_{\max}),\\
&\sum_{J=J_{\min}}^{J_{\max}}\frac{J^2(2 J+1)}{\sqrt{J-J_{\min}}}\mathrm{Bin}(2J_{\max}+1,J_{\max}-J, p)=2c^{\frac{5}{4}}J^{\frac{5}{2}}_{\max}+\mathcal{O}(J^{\frac{3}{2}}_{\max}).
\label{app:unbiased1}
\end{align}
It follows that  
\be
\langle c_{\rm op}^{\rm loc}\rangle=c+\mathcal{O}(J^{-1}_{\max}).
\label{app:unbiased2}
\ee
A similar calculation gives for the MSE of the estimator 
\be
\langle(c_{\rm op}^{\rm loc}-c)^{2}\rangle:=\frac{2c(1-c)}{J_{\max}}+\mathcal{O}\left(J^{-\frac{3}{2}}_{\max}\right).
\label{app:varest}
\ee

\section{Optimal global mean squared error}
\label{glMSE}
In this section we derive the optimal estimator and corresponding global mean squared error (glMSE) for the case where 
the overlap $c$ is a random variable with a distribution induced by the Haar-uniform measure of 
$\mathrm{SU}(d)$. 

The result we obtain is summarized by the following theorem:

\begin{theorem*}
Given $N$ and $M$ copies of two Haar-random states, respectively $\ket{\psi}$ and the state $\ket{\phi}$, define $\ket{\Psi}=\ket{\phi}^{\otimes N}\otimes\ket{\psi}^{\otimes M}$. The global mean square error on the estimation of the overlap $c=|\braket{\psi}{\phi}|^2$ attained by a POVM $\{E_k\}_k$ and an estimator $c_k$ is
\be
v=\sum_{k}\int dU\,dV\,(c(k)-c)^{2} \tr{E_{k}\proj{\Psi}}.
\ee
The minimum of $v$ is attainable by a projective measurement and it reads

\be
v_{\rm op}=\frac{(d-1)(d+M+N)}{d(1+d)(d+M)(d+N)}.
\ee
\end{theorem*}

We follow the standard treatment in~\cite{Personick} to compute the estimator that minimizes glMSE, and use representation theory to perform the integrations.

The probability distribution of the overlap, induced by the Haar measure, is ~\cite[Eq.~(13)]{OverlapStatistics}
\be
p(c)=\int_{\mathrm{SU}(d)}\d U\,\delta(c-|\bra{\psi}U\ket{\psi}|^2)=(d-1)(1-c)^{d-2}
\label{app:haarmeasure}
\ee 

Following~\cite{Personick} the optimal estimator, $S$, satisfies
\be
\frac{S\Gamma+\Gamma S}{2}=\eta
\label{app:optimalS1}
\ee
where
\begin{align}\nonumber
\Gamma&\equiv\int p(c)\, \rho(c)\, \mathrm{d}c\\
\eta&\equiv\int c\, p(c)\, \rho(c)\, \mathrm{d}c.
\label{app:Gammaandeta}
\end{align}
and is explicitly given by 
\be
S=\int_{0}^{\infty} e^{-\alpha\Gamma}\,\eta\, e^{-\alpha\Gamma}\, \d\alpha
\label{app:optimalS}
\ee

Plugging Eq.~\eqref{app:overlap_pure} into Eqs.~(\ref{app:Gammaandeta},\ref{app:optimalS}) gives
\begin{align}\nonumber
S&=\sum_{J=J_{\min}}^{J_{\max}}\frac{\int c\, p(J,c)\d c}{\int p(J,c)\,\d c}\one_{\mathcal U^{(J)}(\mathrm{SU}(d))}\otimes\one_{\mathcal U^{(J)}(S_{N+M})}=\sum_{J=J_{\min}}^{J_{\max}}\frac{\tr{\Pi_{J}\eta}}{\tr{\Pi_{J}\Gamma}}\one_{\mathcal U^{(J)}(\mathrm{SU}(d))}\otimes\one_{\mathcal U^{(J)}(S_{N+M})}\\ \nonumber
&=\sum_{J=J_{\min}}^{J_{\max}}\int c\, p(c|J)\, \d c \,\one_{\mathcal U^{(J)}(\mathrm{SU}(d))}\otimes\one_{\mathcal U^{(J)}(S_{N+M})}\\
&=\sum_{J=J_{\min}}^{J_{\max}}\langle c \rangle_{J}\,\one_{\mathcal U^{(J)}(\mathrm{SU}(d))}\otimes\one_{\mathcal U^{(J)}(S_{N+M})},
\label{app:estimator}
\end{align}
where $p(J,c)=p(c)p(J|c)$, $\int p(J,c)\,\d c=p(J)$ and we have used Bayes' theorem in 
going from the second to the third line of Eq.~\eqref{app:estimator}.  Again, the optimal measurement corresponds to measuring 
the total irrep label $J$.  Upon a given outcome the estimator that minimizes the glMSE is 
$c_{opt}^{bay}(j)=\langle c\rangle_{J}$ where the expectation value is taken with respect to the conditional 
probability distribution $p(c|J)$. 

The operators $\Gamma$ and $\eta$ are given by 
\begin{align}\nonumber
\Gamma&=\int_0^1 p(c) \left(\int_{\mathrm{SU}(d)}\d U\left(U\proj{\psi}U^\dagger\right)^{\otimes N}\otimes\left(UT(c)\proj{\psi}T^\dagger(c) U^\dagger\right)^{\otimes M}\right)\, \d c\\  \nonumber
&=\int_{\mathrm{SU}(d)}\d U\left(U\proj{\psi}U^\dagger\right)^{\otimes N}\otimes \int_0^1 \d c\int_{\mathrm{SU}(d)}\d V\, \delta(c-|\bra{\psi}V\ket{\psi}|^2)  \left(UT(c)\proj{\psi}T^\dagger(c) U^\dagger\right)^{\otimes M}\\  \nonumber
&=\int_{\mathrm{SU}(d)}\d U\left(U\proj{\psi}U^\dagger\right)^{\otimes N}\otimes \int_{\mathrm{SU}(d)}\d V\,\left(UW_{V}V\proj{\psi}U^\dagger(h)W_{V}^\dagger U^\dagger\right)^{\otimes M}\\  \nonumber
&=\int_{\mathrm{SU}(d)}\d U\left(U\proj{\psi}U^\dagger\right)^{\otimes N}\otimes\int_{\mathrm{SU}(d)}\d V\left(V\proj{\psi}U^\dagger(h)\right)^{\otimes M}\\  \nonumber
&=\cG_{\mathrm{SU}(d)}\left[\left(\proj{\psi}\right)^{\otimes N}\right]\otimes \cG_{\mathrm{SU}(d)}\left[\left(\proj{\psi}\right)^{\otimes M}\right]\\ \nonumber
&=\frac{\one_{\mathcal U^{(\frac{N}{2})}(\mathrm{SU}(d))}}{\chi_{\frac N 2}}\otimes \frac{\one_{\mathcal U^{(\frac{M}{2})}(\mathrm{SU}(d))}}{\chi_{\frac M 2}}\\
&=\frac{1}{\chi_{\frac N 2}\chi_{\frac M 2}}\sum_{J=\frac{|N-M|}{2}}^{\frac{N+M}{2}}\one_{\mathcal U^{(J)}(\mathrm{SU}(d))}\otimes\one_{\mathcal U^{(J)}(S_{N+M})},
\label{app:Gamma}
\end{align}
where we have made use of the fact that $W_{V}V=T(c)$ for a unitary $W_{V}$ such that $W_{V}\ket{\psi}\bra{\psi}W_{V}^\dagger =\ket{\psi}\bra{\psi}$ in the third equality, the invariance of the Haar measure for the fourth equality, and used the addition rules for $\mathrm{SU}(d)$ representations for the last equality.
 
To compute $\eta$ we make use of

\begin{multline}
\int_{\mathrm{SU}(2)} \d g (d-1)\left(1-|D^{\left(\tfrac{1}{2}\right)}_{\tfrac 1 2,\tfrac 1 2}(g)|^2\right)^{d-2} |D^{\left(\tfrac{1}{2}\right)}_{\tfrac 1 2,\tfrac 1 2}(g)|^2 \left(\ket{0}\bra{0}\right)^{\otimes N}\otimes\left(D^{\left(\tfrac{1}{2}\right)}(g)^\dagger \ket{0}\bra{0}D^{\left(\tfrac{1}{2}\right)}(g)\right)^{\otimes M}=\\
\int_{\mathrm{SU}(2)} \d g D^{\tfrac{d-1}{2}}_{-\tfrac{d-3}{2},\tfrac{d-1}{2}}(g)D^{\tfrac{d-1}{2}}_{-\tfrac{d-3}{2},\tfrac{d-1}{2}}(g)^*D^{\tfrac{M}{2}}_{k,\tfrac{M}{2}}(g)D^{\tfrac{M}{2}}_{k',\tfrac{M}{2}}(g)^* \ket{\tfrac N 2,\tfrac N 2}\bra{\tfrac N 2,\tfrac N 2}\otimes\ket{\tfrac M 2,k}\bra{\tfrac M 2,k'}=\\
\frac{1}{d+M}\sum_{k=-\frac M 2}^{J-\frac N 2}\left(C^{\frac{d-1+M}{2},-\frac{d-3}{2}+h}_{\frac{d-1}{2},-\frac{d-3}{2},\frac{M}{2},k} C^{{J},\frac{N}{2}+k}_{\frac{N}{2},\frac{N}{2},\frac{M}{2},k} \right)^2 \ket{J,\frac{N}{2}+k}\bra{J,\frac{N}{2}+k},
\label{app:eta}
\end{multline}
with $D^{j}_{m,n}(g)$ being Wigner matrices, so that 
\begin{align}\nonumber
\eta&=\int_0^1 p(c) c \left(\int_{\mathrm{SU}(d)}\d U\left(U\proj{\psi}U^\dagger\right)^{\otimes N}\otimes\left(UT(c)\proj{\psi}T^\dagger(c) U^\dagger\right)^{\otimes M}\right)\, \d c \\  \nonumber
&=\int_{\mathrm{SU}(d)}\d U\left(U\proj{\psi}U^\dagger\right)^{\otimes N}\otimes \int_0^1\d c\, p(c)\, c \int_{\mathrm{SU}(\mathcal E)}\d V \delta(c - |\bra{\psi}V\ket{\psi}|^2) \left(UT(c)\proj{\psi}T^\dagger(c) U^\dagger\right)^{\otimes M}\\ \nonumber
&=\cG_{\mathrm{SU}(d)}\left[\int_{\mathrm{SU}(\mathcal E)}\d V (d-1)(1-|\bra{\psi}V\ket{\psi}|^2)^{d-2} |\bra{\psi}V\ket{\psi}|^2  \proj{\psi}^{\otimes N}\otimes\left(V\proj{\psi}V^\dagger\right)^{\otimes M}\right]\\ \nonumber
&=\frac{1}{d+M}\sum_{J=J_{\min}}^{J_{\max}}\sum_{k=-J-\frac{N}{2}}^{J-\frac{N}{2}}\left(C^{\frac{d-1+M}{2},-\frac{d-3}{2}+k}_{\frac{d-1}{2},-\frac{d-3}{2};\frac{M}{2},k}\,C^{J,\frac{N}{2}+k}_{\frac{N}{2},\frac{N}{2};\frac{M}{2},k}\right)^2\frac{\one_{\mathcal U^{(J)}(\mathrm{SU}(d))}}{\chi_{J}}\otimes \proj{J}_{\mathcal U^{(J)}(S_{N+M})}.
\end{align}

It is now trivial to compute the optimal Bayesian estimator for a given measurement outcome $J$:

\begin{align}\nonumber
c(J)&=\frac{\tr{(\Pi_{J}\eta)}}{\tr{(\Pi_{J}\Gamma)}}=\frac{\frac{\chi_{J}}{\chi_{\frac N 2}\chi_{\frac M 2}}}{\frac{1}{d+M}\sum_{k=-J-\frac{N}{2}}^{J-\frac{N}{2}}\left(C^{\frac{d-1+M}{2},-\frac{d-3}{2}+k}_{\frac{d-1}{2},-\frac{d-3}{2};\frac{M}{2},k}\,C^{J,\frac{N}{2}+k}_{\frac{N}{2},\frac{N}{2};\frac{M}{2},k}\right)^2},\\
\label{app:Bayesianestimator}
\end{align}
$\chi_{J}$ can be calculated with the Hook formula \cite{HayashiGroupRep}. For Young diagrams with two rows one has
\begin{align} \nonumber
&\chi_{J}=(2J+1)\frac{\left(d+J+\frac{N+M}{2}-1\right)!\left(d-J+\frac{N+M}{2}-2\right)!}{(d-1)!(d-2)!\left(\frac{N+M}{2}+J+1\right)!\left(\frac{N+M}{2}-J\right)!}.\\ \nonumber
\end{align}
To simplify the denominator we employ the graphical calculus techniques in ~\cite{Varshalovich1988}:  \begin{align} \nonumber
&\frac{1}{d+M}\sum_{k=-J-\frac{N}{2}}^{J-\frac{N}{2}}\left(C^{\frac{d-1+M}{2},-\frac{d-3}{2}+k}_{\frac{d-1}{2},-\frac{d-3}{2};\frac{M}{2},k}\,C^{J,\frac{N}{2}+k}_{\frac{N}{2},\frac{N}{2};\frac{M}{2},k}\right)^2=
(2J+1)\sum_{L=\frac{d-3+N}{2}}^{\frac{d-1+N}{2}}\left(C^{L,\frac{d-3+N}{2}}_{\frac{d-1}{2},\frac{d-3}{2};\frac{N}{2},\frac{N}{2}}\right)^2\left\{\begin{matrix}\frac{M}{2} && \frac{d-1}{2} && \frac{d+n-1}{2} \\ L && J && \frac{N}{2} \end{matrix}\right\}^2\\  \nonumber
&=(d-1)(2J+1)(4d + 4J + 4J^2 + 2N - N^2 + 2M + 2NM - M^2)\\
&\times\frac{N! M!(d-1+J +\frac{N+M}{2})!(d-2-J +\frac{N+M}{2})!}{4(d+N)!(d+M)!(-J +\frac{N+M}{2})!(1+J +\frac{N+M}{2})!}
\label{app:valuesofsums}
\end{align}
where the term in curly brackets is the Wigner 6-j symbol. Plugging everything together the optimal glMSE estimator for a given measurement outcome $J$ is given by
\be
c_{opt}^{bay}(J)=\frac{d+J+J^{2}+\frac{M+N}{2}-\left(\frac{M+N}{2}\right)^{2}+MN}{(d+M)(d+N)}
\label{app:glMSEE}
\ee
with its corresponding glMSE
\be
v_{\rm op}^{}=\langle(c^{bay}_{op}-c)^2\rangle= \int_0^1 p(c) c^2 - \sum_J p(J) c(J)^2= \frac{(d-1)(d+M+N)}{d(1+d)(d+M)(d+N)}.
\label{app:glMSEeq}
\ee

\section{1-LOCC strategies}
\label{app:EandP}
Here we derive the pointwise MSE attainable by the estimate-and-project (EP) and estimate-and-estimate (EE) strategies, and study the optimal Bayesian estimator for both.
\subsubsection{Estimate-and-project: local estimation}
The EP strategy consists in first estimating at best one of the states using an optimal collective measurement on all of its copies, and then projecting each copy of the other state on the estimate of the first one. The optimal estimation of a random state $\ket\phi$, given $N$ copies of it, is provided by the covariant measurement of Ref.~\cite{Hayashi1997}, $\{M_{dV}\}$, that produces an estimate $\ket{\phi_{V}}=V\ket 0$, $V\in \mathrm{SU}(d)$, with probability density 
\be\label{eq:hayashi}
\d\mu(V)=\chi_{\frac M 2}|\braket{\phi_{V}}{\phi}|^{2M}dV.
\ee
Finally, we perform the projective measurement $\{\dketbra{\phi_{V}},\one-\dketbra{\phi_{V}}\}$ on each copy of $\ket\psi$. This succeeds with probability 
\be
p_{V}(c)=\abs{\braket{\phi_{V}}{\psi}}^{2}.
\ee
The overall measurement operator is then 
$E_{V,k}^{(ep)}=\d V E_{V}^{(M)}\otimes V^{\otimes N}\Pi_{k}^{(N)}V^{\dag\otimes N}
$, as defined in the main text. Its outcome statistics for a fixed value of the overlap can be written as
\be
\small p(k|c)=\chi_{\frac M2}\int_{\mathrm{SU}(d)}\d V\tr{\left(\dketbra{0}^{\otimes M}\otimes\Pi_{k}^{(N)}\right)V^{\otimes (M+N)}\left(\left(U_{c}\dketbra{0}U_{c}^{\dag}\right)^{\otimes M}\otimes\dketbra{0}^{\otimes N}\right)V^{\dag\otimes (M+N)}},\label{pkc}
\ee
where we have set without loss of generality $\ket{\psi}=\ket0$, $\ket{\phi}=U_{c}\ket 0$ thanks to the presence of the 
reference-frame average. We now consider both local and Bayesian estimation with this strategy. 

For local estimation with EP, we employ as an estimator $c_{\rm ep}^{\rm loc}(k)=\frac kN$, i.e., the fraction of successful projections. This estimator is in general biased, except in the limit of large $N$, and it is not necessarily optimal but provides a natural guess for the overlap given our strategy. Its MSE is
\be\begin{aligned}
v_{ep}(c)&=\int_{\mathrm{SU}(d)}\;\d\mu(V) \sum_{k=0}^{N}\mathrm{Bin}\left(k,N,p_{V}(c)\right)\left(\frac{k}{N}-c\right)^2\\
&=\chi_{\frac M 2}\int_{\mathrm{SU}(d)} \d\mu(V) \left(c^{2}-2c p_{V}(c)+\frac 1 N p_{V}(c)(1-p_{V}(c))+p_{V}(c)^{2}\right)\\
&=c^2-\left(2c-\frac 1 N\right)I_{1}(M)+\left(1-\frac 1 N \right)I_{2}(M),\label{app:eProjApp}
\end{aligned}\ee
where in the first equality we have introduced the binomial distribution $\mathrm{Bin}(k,N,p)$ of $m$ successes out of $N$ trials, with a single-trial success probability $p$, in the second equality we expanded the square and used the mean and variance of $\mathrm{Bin}(k,N,p)$,  and in the third equality we defined the integrals 
\be
I_{i}(M)=\chi_{\frac M2}\int_{\mathrm{SU}(d)} dV \abs{\braket{\phi_{V}}{\phi}}^{2M}\abs{\braket{\phi_{V}}{\psi}}^{2i}.
\ee These can be computed by expanding the scalar products and writing the states in a collective-spin basis, obtaining the expectation value of the operator in Eq. \eqref{app:su(d)twirling}, with the substitution $N\mapsto i$:
\be
I_{i}(M)=\chi_{\frac M2}\bra{0}^{\otimes{M+i}}\cG_{\mathrm{SU}(d)}\left[\ket{\phi}\bra{\phi}^M\otimes \ket{\psi}\bra{\psi}^i\right]\ket{0}^{\otimes{M+i}},
\ee
where we have eliminated the integral over $V$ by including this rotation into the group average.
The result is then given by Eqs.~(\ref{app:probJgivenc},\ref{eq:average_state_diag}) with the same substitution. Plugging in the expression of $\chi_{J}=\left(\begin{array}{c}2J+d-1\\d-1\end{array}\right)$, Ref.~\cite{Hayashi1997}, we have
\be\label{integralevery}
I_{i}(M)=\left(\begin{array}{c}M+i+d-1\\i\end{array}\right)^{-1}(1-c)^{i}P_{i}^{(0,M-i)}\left(\frac{1+c}{1-c}\right),
\ee
which can be computed explicitly for $i=1,2$. Inserting these expressions in Eq.~\eqref{app:eProjApp} we finally obtain
\be
\small v_{ep}(c)=\frac{ 
 c^2 (d M + d^2 M + N - 2 M N - N^2) + 
 c (-2 M - 2 d M - 3 N + d N + 2 M N + N^2) + d-1 + 2 M + N}{
 M (d + N) (1 + d + N)}.
\ee
In the limit $M\rightarrow \infty$, $N$ constant we have
\be
v_{ep}(c)\sim \frac{c(1-c)}{N},
\ee
which coincides with the optimal strategy, corresponding to a projection on the known direction of $\ket\phi$.
In the limit $M+N\rightarrow\infty$, $M-N$ fixed we have instead
\be\label{MSEequalEP}
v_{ep}(c)\sim\frac{6c(1-c)}{(M+N)},
\ee 
which is $3/2$ times larger than the optimal strategy. 

\subsubsection{Estimate-and-project: Bayesian estimation}

For Bayesian estimation with EP, as in the previous section, the optimal classical estimator is given by 
\be\label{eq:estCl}
c(k)=\int dc\;c\; p(c|k) =\frac{\tilde{c}(k)}{p(k)},\text{ with } \tilde{c}(k):=\int \d c\;c\; p(c)p(k|c).
\ee
We start by computing the probability distribution of the outcomes, using Eq.~\eqref{pkc}:
\be
\small\begin{aligned}
p&(k)=\int dc\;p(c) p(k|c)\\
&=\chi_{\frac M2} \int_0^1\d c p(c) \int_{\mathrm{SU}(d)}\d U\;\tr{\left(\dketbra{0}^{\otimes M}\otimes\Pi_{k}^{(N)}\right)U^{\otimes (M+N)}\left(\left(U(c)\dketbra{0}U^{\dag}(c)\right)^{\otimes M}\otimes\dketbra{0}^{\otimes N}\right)U^{\dag\otimes (M+N)}}\\
&=\int_{\mathrm{SU}(d)}\d U\tr{\left(\dketbra{0}^{\otimes M}\otimes\Pi_{k}^{(N)}\right)\left(\one^{sym}_{M}\otimes\left(U\dketbra{0}U^{\dag}\right)^{\otimes N}\right)}\\
&=\frac{1}{\chi_{\frac N 2}}\tr{\Pi_{k}^{(N)}\one^{sym}_{N}},\label{pk}
\end{aligned}
\ee
where we performed the average over $c\in\mathrm{SU}(d)$ in the third equality, defining $\one_{M}^{sym}=\one_{\mathcal U^{\left(\frac{M}{2}\right)}(\mathrm{SU}(d))}$ as the projector on the completely symmetric subspace of $M$ qudits, and employed its invariance under $U^{\otimes M}$. We are then left to compute the overlap of $\Pi_{k}^{(N)}$ with $\one^{sym}_{N}$. In order to do so, we recall that the latter can be written as the average of all permutation operators $V(\sigma),\, \sigma\in S_{N}$ of $N$ $d$-level systems. Then the symmetrization of $\Pi_{k}^{(N)}$ gives a trivial binomial factor and we can write
\be
\begin{aligned}
p(k)&=\frac{1}{\chi_{\frac N 2} N!}\left(\begin{array}{c}N\\k\end{array}\right)\sum_{\underline{i}\in\{1,\cdots,d-1\}^{N-k}}\sum_{\sigma}\bra{0}^{\otimes k} \bra{\underline{i}} V(\sigma) \ket{0}^{\otimes k} \ket{\underline{i}}\\
&=\frac{1}{\chi_{\frac N 2} N!}\left(\begin{array}{c}N\\k\end{array}\right)\sum_{\underline{i}\in\{1,\cdots,d-1\}^{N-k}}k!\prod_{j=1}^{d-1}\beta_{j}(\ui)!,\label{trPikIn}
\end{aligned}\ee
where $\beta_{j}(\ui)$ is the number of times the integer $j$ appears in the sequence $\ui$. The sum over $\ui$ can then be broken up into the sum over all partitions of $N-k$ systems in $d-1$ sectors, i.e., the sum over all possible vectors $\underline{\beta}$ of $d-1$ components that add up to $N-k$, times the sum of all permutations of $N-k$ systems which are equal in groups of size $\beta_{j}$. The latter can be carried out immediately since the summand is invariant under permutation of the $j$'s:
\be
p(k)=\frac{1}{\chi_{\frac N 2} N!}\left(\begin{array}{c}N\\k\end{array}\right)\sum_{\underline{\beta}\cdot\underline{1}=N-k}\frac{(N-k)!}{\prod_{j=1}^{d-1}\beta_{j}!}k!\prod_{j=1}^{d-1}\beta_{j}!=\frac{1}{\chi_{\frac N 2}}\left(\begin{array}{c}N-k+d-2\\d-2\end{array}\right),
\ee
where in the last equality we summed over $\underline{\beta}$. For the first moment of the distribution we get, similarly,

\begin{eqnarray}
\small\tilde{c}(k)&=&\chi_{\frac M2} \int_0^1\d c p(c)\int_{\mathrm{SU}(d)}\d U\;\tr{\Big(\dketbra{0}^{\otimes (M+1)}\otimes\Pi_{k}^{(N)}\right)\nonumber\\
&\times& \one\otimes U^{\otimes (M+N)}\left(\left(U(c)\dketbra{0}U(c)\right)^{\otimes (M+1)}\otimes \dketbra{0}^{\otimes N}\Big)\one\otimes U^{\dag\otimes (M+N)}}\label{ckt}\\
&=&\frac{\chi_{\frac M2}}{\chi_{\frac{M+1}{2}}}\int_{\mathrm{SU}(d)}\d U\;\,\tr{\Big(\dketbra{0}^{\otimes (M+1)}\otimes\Pi_{k}^{(N)}\right)U^{\dag}\otimes \one^{\otimes M}\otimes U^{\otimes N}\left(\one_{M+1}^{sym}\otimes \dketbra{0}^{\otimes N}\Big)U\otimes \one^{\otimes M}\otimes U^{\dag\otimes N}},\nonumber
\end{eqnarray}
where we have introduced $U^{\dag}U$ and its conjugate on the additional subsystem, then employed the invariance of $\one_{M+1}^{sym}$ under $U^{\otimes (M+1)}$.
In order to proceed, we first compute the value of the following operator:
\be
\begin{aligned}\label{am}
A_{1}(M)&=\ptr{(1,M)}{\one_{M+1}^{sym}\cdot\dketbra{0}^{\otimes M}}=\frac{1}{(M+1)!}\sum_{\sigma}\sum_{i,j=0}^{d-1}\bra{0}^{\otimes M}\bra{i}V(\sigma)\ket{0}^{\otimes M}\ket{j}\ketbra{i}{j}\\
&=\frac{M!}{(M+1)!}\left(\one+M\dketbra{0}\right),
\end{aligned}\ee
where in the first equality we have taken the partial trace over $M$ subsystems, while in the second one we have written $\one_{M+1}^{sym}$ as an average of permutations, like before, and written the explicit basis representation of the last subsystem. The third equality follows by evaluating the only non-zero elements in the sums: the first term contains all the permutations of $M$ subsystems times the identity on the remaining subsystem; the second term considers the additional permutations in the case $i=j=0$, where the last subsystem can be exchanged with any of the other $M$ subsystems. By substituting this expression in Eq.~\eqref{ckt} we obtain
\be
\tilde{c}(k)=\frac{\chi_{\frac M2}}{\chi_{\frac{M+1}{2}}}\left(\frac{p(k)}{M+1}+\frac{M}{M+1}\tr{\left(\dketbra{0}\otimes\Pi_{k}^{(N)}\right)\frac{\one_{N+1}^{sym}}{\chi_{\frac{N+1}{2}}}}\right),
\ee
where we have used Eq.~\eqref{pk}. We then just need to compute the second term in the sum above, which is very similar to Eq.~\eqref{trPikIn} with the change $\ket{0}^{\otimes k}\mapsto\ket{0}^{\otimes (k+1)}$. We finally obtain
\be
\tilde{c}(k)=\frac{(d-1) (d + N+ k (M+1)) N! (N-k+d-2)!}{(d + M) (N-k)!  (d + N)!}
\ee
and the optimal Bayesian estimator for each $k$ is
\be
c(k)=\frac{d + N+ k (M+1)}{(d + M) (d + N)}.
\ee
The corresponding glMSE is given by
\be
v_{ep}=\int \d c\,p(c) c^{2}-\sum_{k=0}^{N}p(k)c(k)^{2}=\frac{(d-1)((d + M)^2 + (d + 2 M) N)}{d (1 + d) (d + M)^2 (d + N)}.
\ee
In the limit $M\rightarrow \infty$, $N$ constant we have
\be
v_{ep}\sim \frac{(d-1)}{d (d+1) (d + N)},
\ee
which again coincides with the optimal Bayesian strategy.
In the limit $M+N\rightarrow\infty$, $M-N$ fixed we have instead
\be
v_{ep}\sim\frac{6 (d-1)}{d (d+1) (M+N)},
\ee 
which again is $3/2$ times larger than the optimal Bayesian strategy. 
\\

\subsubsection{Estimate-and-estimate: Bayesian estimation}

The EE strategy instead consists in estimating both states with a covariant measurement, hence it is described by overall 
POVM operators
$E_{V,k}^{(ep)}=\d V\d W E_{V}^{(M)}\otimes E_{W}^{(N)}$, as mentioned in the main text.
Its success probability can be written as
\be\begin{aligned}\label{pwc}
p(W|c)&=\d W\;\chi_{\frac M2}\,\chi_{\frac N2}\int_{\mathrm{SU}(d)}\d U\mathrm{Tr}\Big[\left(\dketbra{0}^{\otimes M}\otimes\left(W\dketbra{0}W^{\dag}\right)^{N}\right)\\
&\times U^{\otimes (M+N)}\left(\left(U(c)\dketbra{0}U^{\dag}(c)\right)^{\otimes M}\otimes\dketbra{0}^{\otimes N}\right)U^{\dag\otimes (M+N)}\Big],
\end{aligned}\ee
where again we could include one of the outcomes into the unitary average and, since by redefining $W\mapsto V^{\dag}W$ the dependence on $V$ is a constant, we averaged over $V$ without loss of generality. 

In the Bayesian case for EE we proceed as before and compute first
\be
p(W)=\int dc p(c) p(W|c)=dW\tr{\dketbra{0}^{\otimes (M+N)}\left(\one_{M}^{sym}\otimes\one_{N}^{sym}\right)}=dW,
\ee
then
\be\begin{aligned}
\small\tilde{c}(W)&=\int dc p(c) p(W|c) c=dW\frac{\chi_{\frac M2}\chi_{\frac N2}}{\chi_{\frac{M+1}{2}}}\int_{\mathrm{SU}(d)}dU\mathrm{Tr}\Big[\left(\dketbra{0}^{\otimes (M+1)}\otimes\left(W\dketbra{0}W^{\dag}\right)^{N}\right)\cdot\\
&\cdot \one\otimes U^{\otimes (M+N)}\left(\one_{M+1}^{sym}\otimes\dketbra{0}^{\otimes N}\right)\one\otimes U^{\dag\otimes (M+N)}\Big]\\
&=dW\frac{\chi_{\frac M2}\chi_{\frac N2}}{\chi_{\frac{M+1}{2}}}\int_{\mathrm{SU}(d)}dU\mathrm{Tr}\Big[\left(U\dketbra{0}U^{\dag}\otimes(W\dketbra{0}W^{\dag})^{\otimes N}\right)\cdot\\
&\cdot\left(\frac{\one}{M+1}+\frac{M}{M+1}\dketbra{0}\right)\otimes(U\dketbra{0}U^{\dag})^{\otimes N}\Big]\\
&=dW\frac{\chi_{\frac M2}\chi_{\frac N2}}{\chi_{\frac{M+1}{2}}(M+1)}\left(\frac{1}{\chi_{\frac{N}{2}}}+\frac{M}{\chi_{\frac{N+1}{2}}}\tr{\one_{N+1}^{sym}(W^{\dag}\dketbra{0}W\otimes\dketbra{0}^{\otimes N})}\right)\\
&=dW\frac{\chi_{\frac M2}\chi_{\frac N2}}{\chi_{\frac{M+1}{2}}(M+1)}\left(\frac{1}{\chi_{\frac{N}{2}}}+\frac{M}{\chi_{\frac{N+1}{2}}(N+1)}\left(1+N w\right)\right)=\frac{d + M + N + M N w}{(d + M) (d + N)}.
\end{aligned}\ee
The second equality above comes from averaging over $U_{c}$, the third one from Eq.~\eqref{am} and introducing $U^{\dag}U$ and its conjugate on the additional subsystem, the fourth one from redefining $U\mapsto W^{\dag}U$, switching the operators acting on the $N$ subsystems and averaging over $U$, while the fifth one from applying Eq.~\eqref{am} again and defining $w=\abs{\bra{0}W\ket{0}}^{2}$.

Then the optimal EE Bayesian estimator for each $W$ is simply $c(W)=\tilde{c}(W)/dW$ and the minimum  glMSE attained by it is
\be
v_{ee}=\int \d c \,p(c) \,c^{2}-\int_{\mathrm{SU}(d)}\d W {\tilde{c}(W)}^{2}=\frac{(d-1) (d + M + N) (d^2 + 2 M N + d (M + N))}{d (d+1) (d + 
   M)^2 (d + N)^2},
\ee
where we have carried out the group averages in the usual way:
\be
\int_{\mathrm{SU}(d)} \d W w^{i}=\int_{\mathrm{SU}(d)} \d W \tr{\frac{\one_{i}^{sym}}{\chi_{\frac i2}}\dketbra{0}^{\otimes i}}=\frac{1}{\chi_{\frac i2}}.
\ee
In the limit $M\rightarrow \infty$, $N$ constant we have
\be
v_{ee}\sim \frac{(d-1)(d+2N)}{d (d+1) (d + N)^{2}},
\ee
which is $(d+2N)/(d+N)$ times larger than the optimal Bayesian strategy.
In the limit $M+N\rightarrow\infty$, $M-N$ fixed we have instead
\be
v_{ee}\sim\frac{8 (d-1)}{d (d+1) (M+N)},
\ee 
which is $2$ times larger than the optimal Bayesian strategy. 

\subsubsection{Estimate-and-estimate: local estimation}

Finally, for the local EE estimation the estimator $\tilde{c}_{W}=w=\abs{\bra{0}W\ket{0}}^{2}$ is a natural guess. Its variance can be computed in terms of its first and second moments according to the distribution $p(W|c)$:
\be\label{veelocgen}
v_{ee}(c)=\int_{\mathrm{SU}(d)} p(W|c) (w-c)^{2}=c^{2}-2c\overline{w}+\overline{w^{2}},
\ee
where
\be\begin{aligned}
\overline{w^{i}}=\int p(W|c) w^{i}&=\chi_{\frac M2}\, \chi_{\frac N2}\int_{W,U\in \mathrm{SU}(d)}\d W\d U\mathrm{Tr}\Big[\left(\dketbra{0}^{\otimes M}\otimes\left(W\dketbra{0}W^{\dag}\right)^{\otimes (N+i)}\right)\\
&\times U^{\otimes (M+N)}\otimes\one^{\otimes i}\left(\left(U_{c}\dketbra{0}U_{c}^{\dag}\right)^{\otimes M}\otimes\dketbra{0}^{\otimes (N+i)}\right)U^{\dag\otimes (M+N)}\otimes\one^{\otimes i}\Big]\\
&=\frac{\chi_{\frac M2}\,\chi_{\frac N2}}{\chi_{\frac{N+i}{2}}}\int_{\mathrm{SU}(d)}dU\mathrm{Tr}\Big[\left((U_{c}\dketbra{0}U_{c}^{\dag})^{\otimes M}\otimes\one_{N+i}^{sym}\right)\cdot\\
&\cdot\left(\left(U^{\dag}\dketbra{0}U\right)^{\otimes (M+i)}\otimes\dketbra{0}^{\otimes N}\right)\Big].
\end{aligned}\ee
Then the first moment is straightforward to compute by inserting Eq.~\eqref{am}:
\be\begin{aligned}\label{w1}
\overline{w}&=\frac{\chi_{\frac M2}\, \chi_{\frac N2}}{\chi_{\frac{N+1}{2}}(N+1)}\left(\frac{1}{\chi_{\frac M2}}+\frac{N}{\chi_{\frac{M+1}{2}}}\tr{\one_{M+1}^{sym}\left(\dketbra{0}^{\otimes M}\otimes U_{c}^{\dag}\dketbra{0}U_{c}\right)}\right)\\
&=\frac{\chi_{\frac M2}\chi_{\frac N2}}{\chi_{\frac{N+1}{2}}(N+1)}\left(\frac{1}{\chi_{\frac M2}}+\frac{N}{\chi_{\frac{M+1}{2}}}\frac{1+M c}{1+M}\right).
\end{aligned}\ee
For the second moment we first need to evaluate the following operator:
\be
\begin{aligned}\label{am2}
A_{2}(N)&=\ptr{(1,N)}{\one_{N+2}^{sym}\cdot\dketbra{0}^{\otimes N}}=\frac{1}{(N+2)!}\sum_{\sigma}\sum_{\ui,\uj\in\{0,\cdots,d-1\}^{2}}\bra{0}^{\otimes N}\bra{\ui}V(\sigma)\ket{0}^{\otimes N}\ket{\uj}\ketbra{\ui}{\uj}\\
&=\frac{N!}{(N+2)!}\left[2\one_{2}^{sym}+2N\left(\dketbra{0}\otimes\one+\one\otimes\dketbra{0}\right)+N(N-1)\dketbra{00}\right].
\end{aligned}\ee
As before, the third equality follows by evaluating the only non-zero elements in the sums: the first term contains all the permutations of $N$ subsystems times the identity and the swap on the remaining subsystems, which add up to the projector on the completely symmetric subspace of the two subsystems, 
\be
\one_{2}^{sym}=\frac 12\sum_{(i_{1},i_{2})}\left(\dketbra{i_{1},i_{2}}+\ketbra{i_{1},i_{2}}{i_{2},i_{1}}\right);
\ee
the second term considers the additional permutations in the case $i_{1}=j_{1}=0$ and $i_{2}=j_{2}=0$, where one of the remaining subsystems can be swapped or not with the other, then permuted with any of the other $N$ subsystems; analogously, the third term considers the additional permutations in the case $\ui=\uj=\underline{0}$, where each remaining subsystem can be permuted respectively with $N$ and $N-1$ of the others. Hence the second moment of $w$ can be written as
\be\begin{aligned}\label{w2}
\overline{w^{2}}&=\frac{\chi_{\frac{M}{2}}\,\chi_{\frac{N}{2}}}{\chi_{\frac{N+2}{2}}}\int_{\mathrm{SU}(d)} \d U \tr{\left(U_{c}\dketbra{0}U_{c}^{\dag}\right)^{\otimes M}\left(U^{\dag}\dketbra{0}U\right)^{\otimes M}\otimes A_{2}(N)\left(U^{\dag}\dketbra{0}U\right)^{2}}\\
&=\frac{\chi_{\frac{M}{2}}\,\chi_{\frac{N}{2}}}{\chi_{\frac{N+2}{2}}(N+2)(N+1)}\Bigg(\frac{2}{\chi_{\frac{M}{2}}}+\frac{4N}{\chi_{\frac{M+1}{2}}}\tr{A_{1}(M)U_{c}^{\dag}\dketbra{0}U_{c}}\\
&+\frac{N(N-1)}{\chi_{\frac{M+2}{2}}}\tr{A_{2}(M)\left(U_{c}^{\dag}\dketbra{0}U_{c}\right)^{\otimes 2}}\Bigg)\\
&=\frac{\chi_{\frac{M}{2}}\,\chi_{\frac{N}{2}}}{\chi_{\frac{N+2}{2}}(N+2)(N+1)}\left(\frac{2}{\chi_{\frac{M}{2}}}+\frac{4N(1+M c)}{\chi_{\frac{M+1}{2}}(M+1)}+\frac{N(N-1)(2+4 M c+M(M-1)c^{2})}{\chi_{\frac{M+2}{2}}(M+2)(M+1)}\right).
\end{aligned}\ee
By plugging the expressions of Eqs.~(\ref{w1},\ref{w2}) into Eq.~\eqref{veelocgen} we finally get
\be
v_{ee}(c)=\frac{(d + M + 
     N) ((2 + c (c d-2) (d+1) ) ( d + M+N+1) +  2 c (1-c) M N)}{(d + M) (d + 
     M+1) (d + N) ( d + N+1)}.
\ee
In the limit $M\rightarrow \infty$, $N$ finite we have
\be
v_{ee}(c)\sim \frac{2 + c (d+1) (c d-2) + 2 c N - 2 c^2 N}{(d + N) (d + N+1)}\stackrel{N\gg1}{\sim} \frac{2c(1-c)}{N} + O\left(\frac{1}{N^{2}}\right),
\ee
which is twice as large as the optimal strategy in the leading order of $N$.
In the limit $M+N\rightarrow\infty$, $M-N$ fixed we have instead
\be
v_{ee}(c)\sim\frac{8c(1-c)}{(M+N)},
\ee 
which is $2$ times larger than the optimal strategy.
 
\section{Bayesian estimation using the swap test}\label{app:bayes_swap}
Here we derive the formulas employed for the plot of the swap-test performance in Fig.~3 of the main text. The measurement statistics is given by a binomial distribution $\mathrm{Bin}(k,N,p(c))$, of $k$ events out of $N$, with single-event probability $p(c)=\frac{1+c}{2}$. The corresponding optimal classical Bayesian estimator is given by Eq.~\eqref{eq:estCl}, with $p(k|c)\rightarrow \mathrm{Bin}(k,N,p(c))$. We have
\begin{align}&\begin{aligned}
p(c)=\int dc\;p(c) \mathrm{Bin}(k,N,p(c))=\frac{(d-1) \binom{N}{k} \, _2F_1(1,-k;d-1
   k+N;-1)}{2^{N}(d- k+N-1)},
   \end{aligned}\\
   &\begin{aligned}
\tilde{c}(k)=\int dc\;c\; p(c)\mathrm{Bin}(k,N,p(c))=(d-1) 2^{-N} \binom{N}{k} \, _2\tilde{F}_1(2,-k;d-1
   k+N+1;-1) \Gamma (d-k+N-1),
\end{aligned}\end{align}
where $_2F_1$ and $_2\tilde{F}_1$ are hypergeometric and regularized hypergeometric functions {respectively}. Following the derivations of the previous section, the minimum glMSE attainable with the swap test can be written as
\be
v_{\rm sw}^{\rm bay}=\frac{2}{d(d+1)}-\sum_{k=0}^{N}\frac{(d-1) \binom{N}{k} (d-k+N-1) \,
   _2\tilde{F}_1(2,-k;d-k+N+1;-1){}^2 \Gamma (d-k+N-1)^2}{2^{N}\, _2F_1(1,-k;d-k+N;-1)}.
\ee
 
 
\section{Comparison between Schur transform and swap test with imperfect implementations}\label{tradeoff}
In this section we sketch an evaluation of the effect of imperfect gates on the accuracy of the estimate of the overlap. 
First of all we model the error of each iteration of the swap test as white noise for each iteration: $\mathcal N_{sw}(p(c))=(1-\epsilon_{sw})p(c)+\epsilon_{sw} \frac {1}{2} $, the Fisher information becomes

\be
H(\mathcal N_{sw}(p(c)))=\frac{(1-\epsilon_{sw})^2}{1-c^2(1-\epsilon_{sw})^2},
\ee

For $N$ repetitions, one gets a resulting MSE
\be
v_{sw, noisy}(c)=\frac{1-c^2(1-\epsilon_{sw})^2}{(1-\epsilon_{sw})^2N}.
\ee

We model the noise on the Schur transform measurement outcomes also as mixing with a probability distribution $q(c)$:
$\mathcal N_{sw}(p(J|c))=(1-\epsilon_{Sch})p(J|c)+\epsilon_{Sch}q(c)$, with a probability of mixing that scales exponentially in the number of gates, $1-\epsilon_{Sch}\approx (1-\epsilon)^{g}$, where $\epsilon$ is the error per gate, and $g$ is the total number of gates. 
We recall the joint convexity property of the Fisher Information, coming from its monotonicity:
\be\label{jointconv}
F(\lambda p(c)+(1-\lambda)q(c))\leq \lambda F(p(c))+(1-\lambda)F(q(c)),
\ee

If we assume $q(c)$ to be overlap independent, we obtain the bound
\be
F(\mathcal N_{Sch}(p(J|c)))\leq(1-\epsilon_{Sch})F(p(J|c)),
\ee
so that
\be
v_{Sch, noisy}(c)\geq\frac{2c(1-c)}{(1-\epsilon_{Sch})N}.
\ee
This is a very conservative estimate, as we are assuming we acquiring useful information with exponentially small probability. 
Hence the Swap test outperforms our optimal strategy, based on the Schur transform, when the respective implementation errors satisfy the following relation
\be
\frac{(1-\epsilon_{Sw})^2}{(1-c^2)(1-\epsilon_{Sw}^2)}\geq \frac{1-\epsilon_{Sch}}{2c(1-c)}.
\label{noisecond}
\ee

One can express $\epsilon_{Sch}$ and $\epsilon_{Sw}$ in terms of the error per gate, $\epsilon$, raised to gate complexity of 
their respective circuits.   An intermediate strategy could be to  divide the $N$ copies of both $\ket{\phi}$ and $\ket{\psi}$, into  
$R$ groups of $S$ copies, and perform the optimal measurement on each group, followed by classical post-processing. If $N=M=RS$ and $F(J|c,S)$ is the optimal Fisher information for the case with $M=N=S$ copies, the Cramer-Rao bound reads
\be\label{intermediatecr}
v(c)\geq \frac{1}{R F(J|c,S)}.
\ee
The best option would be to choose $S$ as the highest number of copies such that the architecture can perform the optimal measurement in a sufficiently precise way. On the other hand, if one requires to be in the asymptotic regime of the approximation for $c>c_0$, one can just find the minimum $S$ for which the approximation works, and perform the optimal measurement with $S$ copies $R$ times. The classical post processing will have the optimal asymptotic performance for $c>c_0$. In any case the bound \eqref{intermediatecr} is asymptotically achieved by a maximum likelihood estimator when $R\rightarrow\infty$.

%
%

 
\section{Average post-measurement fidelity}\label{app:post_meas}
Here we compute the average post-measurement fidelity for the optimal strategy and the swap test. For the former we have measurement operators $\{\Pi_J\}_J$, so that Eq. (8) of the main text reads
\be
F_{op}=\sum_{J=J_{\min}}^{J_{\max}}\abs{\bra{\Psi}\Pi_{J}\ket{\Psi}}^{2}=\sum_{J}p(J|c)^{2},
\ee
where we have used the $\mathrm{SU}(d)$-invariance of $\Pi_{J}$.
For the swap test we restrict to $M=N$ as usual and we consider that the measurement is separable and identical on each 
couple of copies. Moreover, the measurement on a single pair of copies is a triplet/singlet projection, which is again $\mathrm{SU}(d)$-invariant, and succeeds/fails with probability $(1\pm c)/2$. Hence we have
\be
F_{sw}=\left[\left(\frac{1+c}{2}\right)^{2}+\left(\frac{1-c}{2}\right)^{2}\right]^{N}=\left(\frac{1+c^{2}}{2}\right)^{N}.
\ee

\section{Estimating the overlap between two arbitrary mixed qubits}
\label{app:mixedstates}
In this appendix we derive the optimal estimator and corresponding mean squared error for the case where we are given $N$ 
and $M$ copies of mixed states.  We shall restrict our attention to qubit mixed states and for ease of notation we shall revert to 
the standard angular momentum notation for irrep labels. 

The mixed states whose overlap we wish to estimate are 
\begin{align}\nonumber
\rho_{\psi}^{\otimes N}(r_{0})&=\left(r_{0}\dketbra{\psi}+(1-r_{0})\frac{\mathbf{\one}}{2}\right)^{\otimes N},\\ 
\rho_{\phi}^{\otimes M}(r_{1})&=\left(r_{1}\dketbra{\phi}+(1-r_{1})\frac{\mathbf{\one}}{2}\right)^{\otimes M},
\label{app:mixedstates}
\end{align}  
where $r_{0(1)}$ denotes the corresponding purity of the states.  Following~\cite{EstimationGRC} the states in 
Eq.~\eqref{app:mixedstates} can be written in the total angular momentum basis, after tracing out multiplicities, as 
\begin{align}\nonumber
\tilde{\rho}_{\psi}^{\otimes N}&=\sum_{J_0=0}^{\frac{M}{2}}p_{J_0}\,\tau^{(0)}_{J_0}(\vec{n}_{0})\\
\tilde{\rho}_{\phi}^{\otimes M}&=\sum_{J_1=0}^{\frac{N}{2}}p_{J_1}\,\tau^{(1)}_{J_1}(\vec{n}_{1}),
\end{align}
where
\begin{align}\nonumber
\tau^{(0)}_{J_0}&=\frac{1}{Z^{(0)}_{J_0}}\sum_{k=-J_0}^{J_0}R_{0}^{k}\proj{J_0,k}\\
\tau^{(1)}_{J_1}&=\frac{1}{Z^{(1)}_{J_1}}\sum_{l=-J_1}^{J_1}R_{1}^{l}\sum_{\alpha,\beta=-J_1}^{J_1}d^{(J_1)}_{\alpha,l}(2\cos^{-1}\sqrt{c})d^{(J_1)}_{l,\beta}(2\cos^{-1}\sqrt{c})\ketbra{J_1,\alpha}{J_1,\beta}
\label{app:taus}
\end{align}
with $R_{i}=\frac{1+r_i}{1-r_i},\, Z^{(i)}_{J_i}=\frac{R_{i}^{J_i+1}-R_{i}^{-J_i}}{R_{i}-1}$, and just as for the case of pure states, 
we have chosen $\vec{n}_0=\vec{z}$ without loss of generality.  Moreover, 
\be
p_{J_0}=\left(\frac{1-r^{2}}{4}\right)^{\frac{N}{2}}\left(\begin{array}{c}N\\ \frac{N}{2}-J_{0}\end{array}\right)\frac{2 J_{0} + 1}{\frac{N}{2} + J_{0} + 1}Z_{J_{0}}
\label{app:mixedprobs}
\ee
and similarly for $p_{J_1}$. Using Eq.~\eqref{supp:su(d)twirl} we obtain 
\be
\rho(c)=\cG_{\mathrm{SU}(2)}\left[\rho_\psi^{\otimes N}\otimes\rho_\phi^{\otimes M}\right]=\sum_J\sum_{J_{0},J_{1}}p_{J_0}p_{J_1}\sum_{k,l}\frac{R_0^k R_1^l}{Z^{(0)}_{J_0}Z^{(1)}_{J_1}}\sum_{\alpha=-J_1}^{J_1}\left(C^{J,k+\alpha}_{J_0,k;J_1,\alpha}d^{(J_1)}_{\alpha,l}(2\cos^{-1}\sqrt{c})\right)^2\,\frac{\one_{\mathcal U^{(J)}(\mathrm{SU}(2))}}{2J+1}\otimes\sigma^{(J_0,J_1)},
\ee
where $\sigma^{(J_0,J_1)}\in\cB(\mathcal U^{(J)}(S_{N+M}))$ and they are orthogonal for different couples $(J_0,J_1)$.
To calculate the glMSE we need to compute the operators $\Gamma, \, \eta$ of Eq.~\eqref{app:Gammaandeta}.  
A similar calculation as in Eq. \eqref{app:Gamma}, \eqref{app:eta} gives
\begin{align}\nonumber
\Gamma&=\cG_{\mathrm{SU}(2)}\left[\rho_\psi^{\otimes N}\right]\otimes\cG_{\mathrm{SU}(2)}\left[\rho_\phi^{\otimes M}\right]\\ \nonumber
&=\left(\sum_{J_0=0}^{\frac{N}{2}}p_{J_0}\frac{\one_{\mathcal U^{(J_0)}(\mathrm{SU}(2))}}{2J_0+1}\otimes\frac{\one_{\mathcal U^{(J)}(S_{M+N})}}{\omega_J}\right)\otimes\left(\sum_{J_1=0}^{\frac{M}{2}} p_{J_1}\frac{\one_{\mathcal U^{(J_1)}(\mathrm{SU}(2))}}{2J_1+1}\otimes \frac{\one_{\mathcal U^{(J)}(S_{M+N})}}{\omega_J}\right)\\
&=\sum_{J_0=0}^{\frac{N}{2}}\sum_{J_1=0}^{\frac{M}{2}}\frac{p_{J_0}p_{J_1}}{(2J_0+1)(2J_1+1)}\sum_{J=\lvert J_0-J_1\rvert}^{J_0+J_1}\one_{\mathcal U^{(J)}(\mathrm{SU}(d))}\otimes\sigma^{(J_0,J_1)},
\label{app:BayesianGamma}
\end{align}
For $\eta$ one obtains
\begin{align}\nonumber
\eta&=\int_{\mathrm{SU}(2)}U^{\otimes(N+M)}\left(\sum_{J_0=0}^{\frac{N}{2}}\sum_{J_1=0}^{\frac{M}{2}}\sum_{k=-J_0}^{J_0}\sum_{l=-J_1}^{J_1}\frac{R_0^kR_1^l}{Z^{(0)}_{J_0}Z^{(1)}_{J_1}}\proj{J_0,k}\right.\\
&\left.\otimes \int_{\mathrm{SU}(2)} \d h \, |D^{\left(\frac 1 2\right)}_{\frac{1}{2}\frac{1}{2}}(h) \rvert^2 D^{(J_1)}(h)\proj{J_1,l}{D^{(J_1)}(h)}^\dagger\right)U^{\dagger\, \otimes (N+M)}.
\end{align}
We finally obtain
\begin{align}\nonumber
\eta&=\sum_{J_0=0}^{\frac{N}{2}}\sum_{J_1=0}^{\frac{M}{2}}\frac{p_{J_0}p_{J_1}}{(2J_0+1)(2J_1+1)}\sum_{J=\lvert J_0-J_1\rvert}^{J_0+J_1}\left(1-\frac{(2J_0+1)(2J_1+1)}{p_{J_0}p_{J_1}}\sum_{k=-J_0}^{J_0}\sum_{l=-J_1}^{J_1}\frac{R_0^kR_1^l}{Z^{(0)}_{J_0}Z^{(1)}_{J_1}}\right.\\  \nonumber
&\left.\sum_{L=\lvert J_1-\frac {1} {2} \rvert}^{J_1+\frac {1}{2}}\sum_{h=-J_1}^{J_1} \frac{\left(C^{L,-\frac{1}{2}+h}_{\frac{1}{2},-\frac{1}{2};J_1,h}C^{L,\frac{1}{2}+l}_{\frac{1}{2},\frac{1}{2};J_1,l}C^{J,k+h}_{J_0,k;J_1,h}\right)^2}{(2L+1)}\right)\frac{\one_{\mathcal U^{(J)}(\mathrm{SU}(2))}}{(2J+1)}\otimes \sigma^{(J_0,J_1)}\\ \nonumber
&=\sum_{J_0=0}^{\frac{N}{2}}\sum_{J_1=0}^{\frac{M}{2}}\frac{p_{J_0}p_{J_1}}{(2J_0+1)(2J_1+1)}\sum_{J=\lvert J_0-J_1\rvert}^{J_0+J_1}\left(1-\frac{(2J_0+1)(2J_1+1)}{p_{J_0}p_{J_1}}\sum_{L=\lvert J_1-\frac {1} {2} \rvert}^{J_1+\frac {1}{2}} \sum_{L'=\lvert J_0-\frac {1} {2} \rvert}^{J_0+\frac {1}{2}}(2J+1)\right.\\
&\left.\times\sum_{k=-J_0}^{J_0}\sum_{l=-J_1}^{J_1}\frac{R_0^k\left(C^{L',\frac{1}{2}+k}_{\frac{1}{2},\frac{1}{2},J_0,k}\right)^2R_1^l\left(C^{L,\frac{1}{2}+l}_{\frac{1}{2},\frac{1}{2},J_1,l}\right)^2}{Z^{(0)}_{J_0}Z^{(1)}_{J_1}} \left\{\begin{matrix}J_1 && \frac{1}{2} && L \\ L' && J && J_0 \end{matrix}\right\}^2\right)\frac{\one_{\mathcal U^{(J)}(\mathrm{SU}(2))}}{(2J+1)}\otimes \sigma^{(J_0, J_1)}.
\end{align}
For a given $J, J_0,\, J_1$ and overlap $c$ the estimator is given by
\be
c(J,J_0,J_1)=\frac{\tr{\Pi_J\left(\Pi_{J_0}\otimes\Pi_{J_1}\eta\right)}}{\tr{\Pi_J\left(\Pi_{J_0}\otimes\Pi_{J_1}\Gamma\right)}}
\ee
and the  glMSE reads
\be
v_{op, mix}= \int_0^1 p(c) c^2 - \sum_{J,J_0,J_1} p(J,J_0,J_1) c(J,J_0,J_1)^2.
\ee

The sums in $L,L',m,k$ can be done exactly. The sums in $J$ at the leading order in $R_0, R_1$ can be done by keeping track of the non-exponentially decaying (in $J$) contributions. The final sum in $J_0$ and $J_1$ can be done using the fact that $p_{J_0}\frac{R_{(0)}^{J_0}}{Z_{J_0}}$ can be written as

\begin{align}
p_{J_0}\frac{R_{(0)}^{J_0}}{Z_{J_0}}&=\left(\frac{1-r^{2}}{4}\right)^{\frac{N}{2}}\left(\begin{array}{c}N\\ \frac{N}{2}-J_{0}\end{array}\right)\frac{2 J_{0} + 1}{\frac{N}{2} + J_{0} + 1} R_{(0)}^{J_0}=\left(\begin{array}{c}N\\ \frac{N}{2}-J_{0}\end{array}\right)\frac{2 J_{0} + 1}{\frac{N}{2} + J_{0} + 1} \left(\frac{1+r_0}{2}\right)^{\frac {N}{2}+J_0} \left(\frac{1-r_0}{2}\right)^{\frac {N}{2}-J_0}\nonumber\\
&=\frac{2 J_{0} + 1}{\frac{N}{2} + J_{0} + 1} \mathrm{Bin}(N,\frac{N}{2}-J_0, \frac{1+r}{2}),
\end{align}
and in the limit $M=\alpha Z, \, N=\beta Z,\, Z\to\infty$ one can approximate the  glMSE expanding in moments around the mean of the binomial distribution. The final result reads
\be
v_{op, mix}=\frac{1}{6Mr_0^2}+\frac{1}{6Nr_1^2}+o(Z^{-1})
\label{app:Bayesianmixedvar}
\ee
in agreement with the pure state case for $d=2, \, r_0=r_1=1$.

 \clearpage
 \end{widetext}


\begin{thebibliography}{99}
\bibitem{OverlapLSB}N. H. Lindner, P. F. Scudo, and D. Bru{\ss}, Int. J. of Quantum Information \textbf{4}, 131 (2006).
\bibitem{Unspeakable} A. Peres and P. F. Scudo, in \textit{Quantum Theory: Reconsideration of Foundations}, edited by A. Khrennikov (V\"{a}xj\"{o} University Press, V\"{a}xj\"{o}  Sweden, 2002), p.283.
\bibitem{ReviewFramesInfo} S. D. Bartlett, T. Rudolph, and R. W. Spekkens, Rev. Mod. Phys. \textbf{79}, 555 (2007).
\bibitem{ResourceAsymmetry} G. Gour,and R. W. Spekkens, New Journal of Physics
\textbf{10} (3), 033023 (2008).
\bibitem{AsymmetryBasic} I. Marvian and R. W. Spekkens, New Journal of
Physics \textbf{15} (3), 033001 (2013).
\bibitem{AsymmetryModes}I. Marvian and R. W. Spekkens, Phys. Rev. A \textbf{90},
062110 (2014).
\bibitem{genSchurEst}I. Marvian and R. W. Spekkens, Communications in Mathematical Physics, 331(2), 431-475 (2014).
\bibitem{ChiribellaMo} Y. Mo and G. Chiribella arxiv preprint, arXiv:1906.01300.
\bibitem{SwapTest}H. Buhrman, R. Cleve, J. Watrous, and R. de Wolf, Phys. Rev. Lett. \textbf{87}, 167902 (2001).
\bibitem{QFinger1} J. N. de Beaudrap, Phys. Rev. A \textbf{69}, 022307 (2004).
\bibitem{QFinger2} N. Kumar, E. Diamanti, and I. Kerenidis, Phys. Rev. A \textbf{95}, 032337 (2017).
\bibitem{FuncOverlap} A. K. Ekert, C. M. Alves, D. K. L. Oi, M. Horodecki, P. Horodecki, and L. C. Kwek, Phys. Rev. Lett. \textbf{88}, 2179011 (2002).
\bibitem{EntDet2} F. Mintert, M. Kus, and A. Buchleitner, Phys. Rev. Lett., \textbf{95}, 260502 (2005).
\bibitem{EntDet3} A. W. Harrow and A. Montanaro, J. ACM \textbf{60}, 3 (2010).
\bibitem{EntDet1} S. P. Walborn, P. H. Souto Ribeiro, L. Davidovich, F. Mintert, and A. Buchleitner, Phys. Rev. A, \textbf{75}, 032338 (2007).
\bibitem{OverlapBRS}S. D. Bartlett, T. Rudolph, and R. W. Spekkens, Phys. Rev. A \textbf{70}, 032321 (2004).
\bibitem{OverlapBIM}E. Bagan, S. Iblisdir, and R. Mu\~{n}oz-Tapia, Phys. Rev. A \textbf{73}, 022341 (2006).
\bibitem{OverlapGI} N. Gisin and S. Iblisdir, European Physical Journal D \textbf{39}, 321-327 (2006).
\bibitem{LearningSwap}L. Cincio, Y. Suba\c{s}{\i}, A. T. Sornborger, and  P. J. Coles, New J. Phys. \textbf{20}, 113022 (2018).
\bibitem{ProgrammableProjective}U. Chabaud, E. Diamanti, D. Markham, E. Kashefi, and A. Joux,  Physical Review A \textbf{98} (6), 062318 (2018).
\bibitem{LloydAlgo}S. Lloyd, M. Mohseni, and P. Rebentrost, arXiv preprint quant-ph/1307.0411 .
\bibitem{QuantumSVM}P. Rebentrost, M. Mohseni, and S. Lloyd, Phys. Rev. Lett. \textbf{113}, 130503 (2014).
\bibitem{HHL} A. W. Harrow, A. Hassidim, and S. Lloyd, Phys. Rev. Lett., \textbf{103}, 150502 (2009).
\bibitem{GaussianProc}Z. Zhao, J.K. Fitzsimons, and J.F. Fitzsimons, Phys. Rev. A \textbf{99}, 052331 (2019).
\bibitem{NNAlgorithm}N. Wiebe, A. Kapoor, and K. Svore, Quantum Inf. Comput. \textbf{15}, 0318-0358 (2015).
\bibitem{IBMSupervised}V. Havlicek, A. D. C\'{o}rcoles, K. Temme, A. W. Harrow, A. Kandala, J. M. Chow, and J. M. Gambetta, Nature \textbf{567}, 209-212 (2019).
\bibitem{Nana} N. Liu and P. Rebentrost, Phys. Rev. A \textbf{97}, 042315 (2018).
\bibitem{variationalnear} H.-Y. Huang, K. Bharti, and Patrick Rebentrost, arXiv preprint arXiv:1909.07344.
\bibitem{SentisUnsuper} G. Sent\'{i}s, A Monr\`{a}s,  R. Mu\~{n}oz-Tapia, and J. Calsamiglia Phys. Rev. X 9, 041029 (2019).
\bibitem{HelstromBOOK} C. W. Helstrom, Quantum detection and estimation theory. Journal of Statistical Physics, 1(2), 231-252 (1969). 
\bibitem{HolevoBook} A. S. Holevo, \textit{Probabilistic and Statistical Aspects of Quantum Theory} (North-Holland, Amsterdam, 1982).
\bibitem{FisherParis} M. G. A. Paris, Int. J. Quant. Inf. \textbf{7}, 125-137 (2009).
\bibitem{Personick} S. D. Personick, IEEE Trans. Inf. Theory \textbf{17}, 240 (1971).
\bibitem{Szego1939}G. Szego, \textit{Orthogonal Polynomials}, 4th ed. Colloquium Publications Vol. 23 (American Mathematical Society, Providence, 1975).
\bibitem{OverlapStatistics}L. Alonso and T. Gorin, Journal of Physics A: Mathematical and Theoretical \textbf{49}, 145004 (2016).
\bibitem{Hayashi1997} M. Hayashi, J. Phys. A. Math. Gen. \textbf{31}, 4633 (1998).
\bibitem{HayashiGroupRep} M. Hayashi, \textit{Group Representation for Quantum Theory}. Berlin: (Springer, Berlin, 2017).

\bibitem{EstimationGRC} B. Gendra, E. Ronco-Bonvehi, J. Calsamiglia, R. Munoz-Tapia, and E. Bagan, New Journal of Physics \textbf{14}, 105015 (2012).

 \bibitem{Varshalovich1988}D. A. Varshalovich, A. N. Moskalev, and V. K. Khersonskii, \textit{Quantum Theory of Angular Momentum}, (World Scientific, Singapore, 1988).

\bibitem{googleplateau}J. R. McClean, S. Boixo, V. N. Smelyanskiy,
R. Babbush, and H. Neven, Nature Communications \textbf{9}, 4812 (2018).
\bibitem{glassyplateau}A. G. R. Day, M. Bukov, P. Weinberg, P. Mehta, and D. Sels, Physical Review Letters \textbf{122}, 020601 (2019).
\bibitem{plateaulearning}X. Glorot and Y. Bengio, in \textit{Proceedings of the 13th International Conference on Artificial Intelligence and Statistics, Sardinia, Italy, 2010}, p. 249.
\bibitem{quantumplateau}M. Benedetti, D. Garcia-Pintos, O. Perdomo, V. Leyton-Ortega, Y. Nam, and A. Perdomo Ortiz, npj Quantum Information \textbf{5}, 45 (2019).
\bibitem{unitaryoverlap} S. Khatri, R. LaRose, A. Poremba, L. Cincio, A. T. Sornborger, and P. J. Coles, Quantum \textbf{3}, 140 (2019).
 \bibitem{Schur1} D. Bacon, I. L. Chuang, and A. W. Harrow, \textit{Proceedings of the Eighteenth Annual ACM-SIAM Symposium on Discrete Algorithms, (SODA '07)}, (Society for Industrial and Applied Mathematics, Philadelphia, 2007). 
 \bibitem{Schur2} H. Krovi, Quantum \textbf{3}, 122 (2019).
 \bibitem{Schur3} A. W. Harrow, Ph.D. thesis, (Massachusetts Institute of Technology, 2005). arXiv preprint: quant-ph/0512255.
\bibitem{BOW} C. B\u{a}descu, R. O'Donnell, and J. Wright, in \textit{Proceedings of the 51st Annual ACM SIGACT Symposium on Theory of Computing}, pp. 503-514 (2019).
\bibitem{Weyl}H. Weyl, \textit{The classical groups: their invariants and representations}, (Princeton University Press, 1949).
\bibitem{Sternberg}S. Sternberg, \textit{Group theory and physics}, (Cambridge University Press, 1995).
\bibitem{Darboux1878}G. Darboux, Journal de Mathématiques Pures et Appliquées (1878): 5-56.
\bibitem{Sanov1957}I. N. Sanov, Mat. Sbornik \textbf{42}, 11-44 (1957) (in Russian). English translation: Selected Translat. Math. Stat. \textbf{1}, 213-244 (1961).

 \end{thebibliography}
 \end{document}